\documentclass[aps,prd,twocolumn,superscriptaddress,showpacs,preprintnumbers,amsmath,amssymb]{revtex4-1}

\usepackage{graphicx}% Include figure files
\usepackage{dcolumn}% Align table columns on decimal point
\usepackage{bm}% bold math
\usepackage[mathlines]{lineno}% Enable numbering of text and display math
\usepackage{subfigure}
\usepackage{multirow}

\begin{document}

\preprint{\vbox{ \hbox{   }
			      \hbox{Belle Preprint 2016-8}
			      \hbox{KEK Preprint 2016-8}
%                  \hbox{Moriond EW 2016}
%                  \hbox{Belle-CONF-1602}
%                  \hbox{Intended for {\it PRL}}
}}

\title{Measurement of the branching ratio of $\bar{B}^0 \rightarrow D^{*+} \tau^- \bar{\nu}_{\tau}$
relative to $\bar{B}^0 \rightarrow D^{*+} \ell^- \bar{\nu}_{\ell}$ decays
with a semileptonic tagging method
}


%%% Paper:    sin(2phi1) in tagged B decays @ Y(5S)
%%% Journal:  Physical Review Letters
%%% Contacts: Y. Sato (yutaro@epx.phys.tohoku.ac.jp)
%%%           H. Yamamoto (yhitoshi@epx.phys.tohoku.ac.jp)
%%% Non-responding authors or those who said NO are commented out.
%%% ====================================================================
%%% Click the RELOAD button on your web browser to see the updated file.
%%% ====================================================================
%%% Use \input{author} to insert this material into your latex file.
%%%%% Force institutions to appear in alphabetical order when typeset.

\date{\today}
%%% Paper:
%%% Journal:  Physical Review
%%% Contacts:
%%% Non-responding authors or those who said NO are commented out.
%%% ====================================================================
%%% Click the RELOAD button on your web browser to see the updated file.
%%% ====================================================================
%%% Use \input{author} to insert this material into your latex file.
%%%%% Force institutions to appear in alphabetical order when typeset.
\noaffiliation
\affiliation{Aligarh Muslim University, Aligarh 202002}
\affiliation{University of the Basque Country UPV/EHU, 48080 Bilbao}
\affiliation{Beihang University, Beijing 100191}
\affiliation{University of Bonn, 53115 Bonn}
\affiliation{Budker Institute of Nuclear Physics SB RAS, Novosibirsk 630090}
\affiliation{Faculty of Mathematics and Physics, Charles University, 121 16 Prague}
%%%\affiliation{Chiba University, Chiba 263-8522}
%%%\affiliation{Chonnam National University, Kwangju 660-701}
\affiliation{University of Cincinnati, Cincinnati, Ohio 45221}
\affiliation{Deutsches Elektronen--Synchrotron, 22607 Hamburg}
\affiliation{University of Florida, Gainesville, Florida 32611}
%%%\affiliation{Department of Physics, Fu Jen Catholic University, Taipei 24205}
\affiliation{Justus-Liebig-Universit\"at Gie\ss{}en, 35392 Gie\ss{}en}
%%%\affiliation{Gifu University, Gifu 501-1193}
%%%\affiliation{II. Physikalisches Institut, Georg-August-Universit\"at G\"ottingen, 37073 G\"ottingen}
\affiliation{SOKENDAI (The Graduate University for Advanced Studies), Hayama 240-0193}
%%%\affiliation{Gyeongsang National University, Chinju 660-701}
\affiliation{Hanyang University, Seoul 133-791}
\affiliation{University of Hawaii, Honolulu, Hawaii 96822}
\affiliation{High Energy Accelerator Research Organization (KEK), Tsukuba 305-0801}
\affiliation{J-PARC Branch, KEK Theory Center, High Energy Accelerator Research Organization (KEK), Tsukuba 305-0801}
%%%\affiliation{Hiroshima Institute of Technology, Hiroshima 731-5193}
\affiliation{IKERBASQUE, Basque Foundation for Science, 48013 Bilbao}
%%%\affiliation{University of Illinois at Urbana-Champaign, Urbana, Illinois 61801}
\affiliation{Indian Institute of Science Education and Research Mohali, SAS Nagar, 140306}
\affiliation{Indian Institute of Technology Bhubaneswar, Satya Nagar 751007}
\affiliation{Indian Institute of Technology Guwahati, Assam 781039}
\affiliation{Indian Institute of Technology Madras, Chennai 600036}
%%%\affiliation{Indiana University, Bloomington, Indiana 47408}
\affiliation{Institute of High Energy Physics, Chinese Academy of Sciences, Beijing 100049}
\affiliation{Institute of High Energy Physics, Vienna 1050}
%%%\affiliation{Institute for High Energy Physics, Protvino 142281}
%%%\affiliation{Institute of Mathematical Sciences, Chennai 600113}
\affiliation{INFN - Sezione di Torino, 10125 Torino}
%%%\affiliation{Advanced Science Research Center, Japan Atomic Energy Agency, Naka 319-1195}
\affiliation{J. Stefan Institute, 1000 Ljubljana}
\affiliation{Kanagawa University, Yokohama 221-8686}
\affiliation{Institut f\"ur Experimentelle Kernphysik, Karlsruher Institut f\"ur Technologie, 76131 Karlsruhe}
%%%\affiliation{Kavli Institute for the Physics and Mathematics of the Universe (WPI), University of Tokyo, Kashiwa 277-8583}
\affiliation{Kennesaw State University, Kennesaw, Georgia 30144}
\affiliation{King Abdulaziz City for Science and Technology, Riyadh 11442}
%%%\affiliation{Department of Physics, Faculty of Science, King Abdulaziz University, Jeddah 21589}
\affiliation{Korea Institute of Science and Technology Information, Daejeon 305-806}
\affiliation{Korea University, Seoul 136-713}
%%%\affiliation{Kyoto University, Kyoto 606-8502}
\affiliation{Kyungpook National University, Daegu 702-701}
\affiliation{\'Ecole Polytechnique F\'ed\'erale de Lausanne (EPFL), Lausanne 1015}
\affiliation{P.N. Lebedev Physical Institute of the Russian Academy of Sciences, Moscow 119991}
\affiliation{Faculty of Mathematics and Physics, University of Ljubljana, 1000 Ljubljana}
\affiliation{Ludwig Maximilians University, 80539 Munich}
\affiliation{Luther College, Decorah, Iowa 52101}
\affiliation{University of Maribor, 2000 Maribor}
\affiliation{Max-Planck-Institut f\"ur Physik, 80805 M\"unchen}
\affiliation{School of Physics, University of Melbourne, Victoria 3010}
\affiliation{Middle East Technical University, 06531 Ankara}
\affiliation{University of Miyazaki, Miyazaki 889-2192}
\affiliation{Moscow Physical Engineering Institute, Moscow 115409}
\affiliation{Moscow Institute of Physics and Technology, Moscow Region 141700}
\affiliation{Graduate School of Science, Nagoya University, Nagoya 464-8602}
\affiliation{Kobayashi-Maskawa Institute, Nagoya University, Nagoya 464-8602}
%%%\affiliation{Nara University of Education, Nara 630-8528}
\affiliation{Nara Women's University, Nara 630-8506}
\affiliation{National Central University, Chung-li 32054}
\affiliation{National United University, Miao Li 36003}
\affiliation{Department of Physics, National Taiwan University, Taipei 10617}
\affiliation{H. Niewodniczanski Institute of Nuclear Physics, Krakow 31-342}
\affiliation{Nippon Dental University, Niigata 951-8580}
\affiliation{Niigata University, Niigata 950-2181}
\affiliation{University of Nova Gorica, 5000 Nova Gorica}
\affiliation{Novosibirsk State University, Novosibirsk 630090}
\affiliation{Osaka City University, Osaka 558-8585}
%%%\affiliation{Osaka University, Osaka 565-0871}
\affiliation{Pacific Northwest National Laboratory, Richland, Washington 99352}
%%%\affiliation{Panjab University, Chandigarh 160014}
%%%\affiliation{Peking University, Beijing 100871}
\affiliation{University of Pittsburgh, Pittsburgh, Pennsylvania 15260}
\affiliation{Punjab Agricultural University, Ludhiana 141004}
%%%\affiliation{Research Center for Electron Photon Science, Tohoku University, Sendai 980-8578}
%%%\affiliation{Research Center for Nuclear Physics, Osaka University, Osaka 567-0047}
\affiliation{Theoretical Research Division, Nishina Center, RIKEN, Saitama 351-0198}
%%%\affiliation{RIKEN BNL Research Center, Upton, New York 11973}
%%%\affiliation{Saga University, Saga 840-8502}
\affiliation{University of Science and Technology of China, Hefei 230026}
\affiliation{Seoul National University, Seoul 151-742}
%%%\affiliation{Shinshu University, Nagano 390-8621}
\affiliation{Showa Pharmaceutical University, Tokyo 194-8543}
\affiliation{Soongsil University, Seoul 156-743}
\affiliation{University of South Carolina, Columbia, South Carolina 29208}
%%%\affiliation{Stefan Meyer Institute for Subatomic Physics, Vienna 1090}
\affiliation{Sungkyunkwan University, Suwon 440-746}
\affiliation{School of Physics, University of Sydney, New South Wales 2006}
\affiliation{Department of Physics, Faculty of Science, University of Tabuk, Tabuk 71451}
\affiliation{Tata Institute of Fundamental Research, Mumbai 400005}
\affiliation{Excellence Cluster Universe, Technische Universit\"at M\"unchen, 85748 Garching}
\affiliation{Department of Physics, Technische Universit\"at M\"unchen, 85748 Garching}
\affiliation{Toho University, Funabashi 274-8510}
%%%\affiliation{Tohoku Gakuin University, Tagajo 985-8537}
\affiliation{Department of Physics, Tohoku University, Sendai 980-8578}
\affiliation{Earthquake Research Institute, University of Tokyo, Tokyo 113-0032}
\affiliation{Department of Physics, University of Tokyo, Tokyo 113-0033}
\affiliation{Tokyo Institute of Technology, Tokyo 152-8550}
\affiliation{Tokyo Metropolitan University, Tokyo 192-0397}
%%%\affiliation{Tokyo University of Agriculture and Technology, Tokyo 184-8588}
\affiliation{University of Torino, 10124 Torino}
%%%\affiliation{Toyama National College of Maritime Technology, Toyama 933-0293}
%%%\affiliation{Utkal University, Bhubaneswar 751004}
\affiliation{Virginia Polytechnic Institute and State University, Blacksburg, Virginia 24061}
\affiliation{Wayne State University, Detroit, Michigan 48202}
\affiliation{Yamagata University, Yamagata 990-8560}
\affiliation{Yonsei University, Seoul 120-749}
  \author{Y.~Sato}\affiliation{Kobayashi-Maskawa Institute, Nagoya University, Nagoya 464-8602}\affiliation{High Energy Accelerator Research Organization (KEK), Tsukuba 305-0801} % Nagoya, KEK
  \author{T.~Iijima}\affiliation{Kobayashi-Maskawa Institute, Nagoya University, Nagoya 464-8602}\affiliation{Graduate School of Science, Nagoya University, Nagoya 464-8602} % Nagoya
% \author{A.~Abdesselam}\affiliation{Department of Physics, Faculty of Science, University of Tabuk, Tabuk 71451} % Tabuk
% \author{I.~Adachi}\affiliation{High Energy Accelerator Research Organization (KEK), Tsukuba 305-0801}\affiliation{SOKENDAI (The Graduate University for Advanced Studies), Hayama 240-0193} % KEK
  \author{K.~Adamczyk}\affiliation{H. Niewodniczanski Institute of Nuclear Physics, Krakow 31-342} % Krakow
  \author{H.~Aihara}\affiliation{Department of Physics, University of Tokyo, Tokyo 113-0033} % Tokyo
% \author{S.~Al~Said}\affiliation{Department of Physics, Faculty of Science, University of Tabuk, Tabuk 71451}\affiliation{Department of Physics, Faculty of Science, King Abdulaziz University, Jeddah 21589} % Tabuk
% \author{K.~Arinstein}\affiliation{Budker Institute of Nuclear Physics SB RAS, Novosibirsk 630090}\affiliation{Novosibirsk State University, Novosibirsk 630090} % BINP
% \author{Y.~Arita}\affiliation{Graduate School of Science, Nagoya University, Nagoya 464-8602} % Nagoya
  \author{D.~M.~Asner}\affiliation{Pacific Northwest National Laboratory, Richland, Washington 99352} % PNNL
% \author{T.~Aso}\affiliation{Toyama National College of Maritime Technology, Toyama 933-0293} % Toyama
  \author{H.~Atmacan}\affiliation{Middle East Technical University, 06531 Ankara} % METU
% \author{V.~Aulchenko}\affiliation{Budker Institute of Nuclear Physics SB RAS, Novosibirsk 630090}\affiliation{Novosibirsk State University, Novosibirsk 630090} % BINP
 \author{T.~Aushev}\affiliation{Moscow Institute of Physics and Technology, Moscow Region 141700} % MIPT
  \author{R.~Ayad}\affiliation{Department of Physics, Faculty of Science, University of Tabuk, Tabuk 71451} % Tabuk
  \author{T.~Aziz}\affiliation{Tata Institute of Fundamental Research, Mumbai 400005} % Tata
  \author{V.~Babu}\affiliation{Tata Institute of Fundamental Research, Mumbai 400005} % Tata
% ...
% Ljubljana
% \author{N.~Zwahlen}\affiliation{\'Ecole Polytechnique F\'ed\'erale de Lausanne (EPFL), Lausanne 1015} % Lausanne
% \author{O.~Zyukova}\affiliation{Budker Institute of Nuclear Physics SB RAS, Novosibirsk 630090}\affiliation{Novosibirsk State University, Novosibirsk 630090} % BINP
\collaboration{The Belle Collaboration}

\begin{abstract}
We report a measurement of the ratio
${\cal R}(D^*) = {\cal B}(\bar{B}^0 \rightarrow D^{*+} \tau^- \bar{\nu}_{\tau})/{\cal B}(\bar{B}^0 \rightarrow D^{*+} \ell^- \bar{\nu}_{\ell})$,
where $\ell$ denotes an electron or a muon.
The results are based on a data sample
containing $772\times10^6$ $B\bar{B}$ pairs
recorded at the $\Upsilon(4S)$ resonance
with the Belle detector at the KEKB $e^+ e^-$ collider.
We select a sample of $B^0 \bar{B}^0$ pairs by reconstructing both $B$ mesons in semileptonic decays to
$D^{*\mp} \ell^{\pm}$.
We measure ${\cal R}(D^*)= 0.302 \pm 0.030({\rm stat)} \pm 0.011({\rm syst)}$,
which is within $1.6 \sigma$ of the Standard Model theoretical expectation,
where the standard deviation $\sigma$ includes systematic uncertainties.
We use this measurement to constrain several scenarios of new physics in a model-independent approach.
\end{abstract}

%\pacs{11.30.Er, 11.30.Hv, 12.15.Ji, 13.20.He}% PACS, the Physics and Astronomy
%\keywords{}

\maketitle

\section{INTRODUCTION}
Semitauonic $B$ meson decays of the type $b \rightarrow c \tau \nu_{\tau}$~\cite{CHARGE_CONJUGATION}
are sensitive probes to search for physics beyond the Standard Model (SM).
Charged Higgs bosons, which appear in supersymmetry~\cite{CHARGED_HIGGS_SUSY}
and other models with at least two Higgs doublets~\cite{CHARGED_HIGGS_2HDM},
may contribute measurably to the decays due to the large mass of the $\tau$.
Similarly, leptoquarks~\cite{LQ1}, which carry both baryon number and lepton number, may also contribute to this process.
The ratio of branching fractions
\begin{eqnarray}
{\cal R}(D^{(*)}) = \frac{{\cal B}(\bar{B} \rightarrow D^{(*)} \tau^- \bar{\nu}_{\tau})}{{\cal B}(\bar{B} \rightarrow D^{(*)} \ell^- \bar{\nu}_{\ell})} \hspace{0.8em}(\ell = e,\mu),
\end{eqnarray}
is typically used instead of the absolute branching fraction of $\bar{B} \rightarrow D^{(*)} \tau^- \bar{\nu}_{\tau}$
to reduce several systematic uncertainties,
such as those on the experimental efficiency, the magnitude of the Cabibbo-Kobayashi-Maskawa matrix element $|V_{cb}|$
and the semileptonic decay form factors.
The SM calculations on these ratios predict
${\cal R}(D^*) = 0.252 \pm 0.003$~\cite{SM_PREDICTION_2} and
${\cal R}(D)   = 0.297 \pm 0.017$~\cite{SM_PREDICTION_1,BABAR_HAD_NEW}
with a precision of better than 2\% and 6\% for ${\cal R}(D^*)$ and ${\cal R}(D)$, respectively.
Consistent values of ${\cal R}(D)$ are predicted using lattice quantum chromodynamics (QCD) calculations:
${\cal R}(D) = 0.299 \pm 0.011$~\cite{SM_PREDICTION_LQCD_1} and
${\cal R}(D) = 0.300 \pm 0.008$~\cite{SM_PREDICTION_LQCD_2}.
Exclusive semitauonic $B$ decays were first observed by Belle~\cite{BELLE_INCLUSIVE_OBSERVATION},
with subsequent studies reported by Belle \cite{BELLE_INCLUSIVE,BELLE_HAD_NEW},
\mbox{\sl B\hspace{-0.4em} {\small\sl A}\hspace{-0.37em} \sl B\hspace{-0.4em}
{\small\sl A\hspace{-0.02em}R}} \cite{BABAR_HAD_NEW},
and LHCb \cite{LHCB_RESULT}.
All the experimental results are consistent with each other,
and the average values of Refs.~\cite{BELLE_HAD_NEW,BABAR_HAD_NEW,LHCB_RESULT} are
${\cal R}(D^*) = 0.322 \pm 0.018 \pm 0.012$
and 
${\cal R}(D) = 0.391 \pm 0.041 \pm 0.028$~\cite{HFAG},
which exceed the SM predictions
by $3.0 \sigma$ and $1.7 \sigma$, respectively.
The combined analysis of ${\cal R}(D^*)$ and ${\cal R}(D)$, taking into account measurement correlations, finds that the deviation is $3.9\sigma$ from the SM prediction~\cite{HFAG}.

So far,  measurements of ${\cal R}(D^{(*)})$ at the $B$ factories
have been performed using hadronic \cite{BELLE_HAD_NEW,BABAR_HAD_NEW} or inclusive tagging methods \cite{BELLE_INCLUSIVE_OBSERVATION,BELLE_INCLUSIVE}.
Semileptonic tagging methods have been employed in studies of $B^- \rightarrow \tau^- \bar{\nu}_{\tau}$ decays
and have been shown to have similar experimental precision
to that of the hadronic tagging method \cite{TAUNU_BELLE_SEMILEP,TAUNU_BABAR_SEMILEP}.
In this paper, we report the first measurement of ${\cal R}(D^*)$ using the semileptonic tagging method.
We reconstruct signal $B^0\bar{B}^0$ events in modes where one $B$ decays semitauonically,
$\bar{B}^0 \rightarrow D^{*+} \tau^- \bar{\nu}_{\tau}$
followed by $\tau^- \rightarrow \ell^- \bar{\nu}_{\ell} \nu_{\tau}$ 
(referred to hereinafter as $B_{\rm sig}$),
and the other $B$ decays in a semileptonic channel $\bar{B}^0 \rightarrow D^{*+} \ell^- \bar{\nu}_{\ell}$
(referred to hereinafter as $B_{\rm tag}$).
In order to form ${\cal R}(D^*)$,
we also reconstruct normalization $B^0\bar{B}^0$ events
in modes where both $B$ mesons decay to
$D^{*+} \ell^- \bar{\nu}_{\ell}$.

\section{DETECTOR AND MC SIMULATION}
We use the full $\Upsilon(4S)$ data sample
containing $772 \times 10^6$ $B \bar{B}$ pairs
recorded with the Belle detector \cite{BELLE}
at the KEKB $e^+ e^-$ collider \cite{KEKB}.
The Belle detector is a general-purpose magnetic spectrometer,
which consists of
a silicon vertex detector (SVD),
a 50-layer central drift chamber (CDC),
an array of aerogel threshold Cherenkov counters (ACC),
time-of-flight scintillation counters (TOF),
and an electromagnetic calorimeter (ECL) comprising CsI(Tl) crystals.
The devices are located inside a superconducting solenoid coil
that provides a 1.5 T magnetic field.
An iron flux-return yoke located outside the coil
is instrumented to detect $K_L^0$ mesons
and to identify muons (KLM).
The detector is described in detail elsewhere \cite{BELLE}.
 
To determine the reconstruction efficiency and probability density functions (PDF)  for
signal, normalization, and background processes,
we use Monte Carlo (MC) simulated events, which are generated with the EvtGen event generator \cite{EVTGEN}
and simulated with the GEANT3 package \cite{GEANT}.
The MC samples for signal events
are generated using the decay model based on the heavy quark effective theory (HQET) in Ref.~\cite{SIG_DECAY_MODEL}.
The normalization mode is simulated using HQET, and reweighted
according to the current world-average form factor values: $\rho^2 = 1.207 \pm 0.015 \pm 0.021$, $R_1 = 1.403 \pm 0.033$,
and $R_2 = 0.854 \pm 0.020$ \cite{HFAG}.
Background $B \rightarrow D^{**} \ell \nu_{\ell}$ events are simulated
with the ISGW \cite{ISGW} model and reweighted to match
the kinematics predicted by the LLSW model \cite{LLSW}.
Here, $D^{**}$ denotes the orbitally excited states $D_1$, $D_2^*$, $D_1'$, and $D_0^*$.
Radially excited states are neglected.
We consider $D^{**}$ decays to a $D^{(*)}$ and a pion, a $\rho$ or an $\eta$ meson,
or a pair of pions,
where branching fractions are assumed
based on quantum-number, phase-space, and isospin arguments.
The sample sizes of the signal, $B\bar{B}$, and continuum $q\bar{q}$ $(q=u,d,s,c)$ production processes
correspond to about 40, 10, and 6 times the integrated luminosity of the on-resonance $\Upsilon(4S)$ data sample, respectively.
% D** lnu (decay model, composition ?)
% D* lnu (form factor ?)

\section{EVENT SELECTION}
%\subsection{charged tracks}
Charged particle tracks are reconstructed with the SVD and CDC.
All tracks other than $K_S^0 \rightarrow \pi^+ \pi^-$ decay daughters are required to originate from near the interaction point (IP).
Electrons are identified by a combination of the specific ionization ($dE/dx$) in the CDC,
the ratio of the cluster energy in the ECL
to the track momentum measured with the SVD and CDC,
the response of the ACC,
the shower shape in the ECL,
and the match between the positions of the shower and the track at the ECL~\cite{EID}.
To recover bremsstrahlung photons from electrons,
we add the four-momentum of each photon detected
within 0.05 radians of the original track direction
to the electron momentum.
Muons are identified
by the track penetration depth and hit distribution in the KLM~\cite{MUID}.
Charged kaons are identified
by combining information from
the $dE/dx$ in the CDC, 
the flight time measured with the TOF,
and the response of the ACC \cite{PID}.
We do not apply any particle identification criteria on charged pions.

Candidate $K_S^0$ mesons are formed
by combining two oppositely charged tracks
with pion mass hypotheses.
We require their invariant mass to lie within 15 MeV/$c^2$ of the nominal $K^0$ mass \cite{PDG},
which corresponds to approximately $7\sigma$,
where $\sigma$ denotes the resolution of the $\pi^+ \pi^-$ invariant mass.
We then impose the following additional requirements:
both pion tracks must have a large distance of closest approach to the IP
in the plane perpendicular to the electron beam line;
the pion tracks must intersect at a common vertex
that is displaced from the IP; and
the momentum vector of the $K_S^0$ candidate should originate from the IP. 

Neutral pion candidates are formed from pairs of photons
with further criteria specific to whether the $\pi^0$ is from a $D^{*+}$ decay or $D$ decay.
For neutral pions from $D$ decays,
we require 
the photon daughter energies to be greater than 50 MeV,
the cosine of the angle in the laboratory frame
between the two photons to be greater than zero,
and the $\gamma \gamma$ invariant mass to be within $-15$ and $+10$ MeV/$c^2$ of the nominal $\pi^0$ mass \cite{PDG},
which corresponds to approximately $\pm 1.8 \sigma$.
Photons are measured as an energy cluster
in the ECL with no associated charged tracks.
A mass-constrained fit is then performed to obtain the $\pi^0$ momentum.
For neutral pions from $D^{*+}$ decays, which have lower energies,
we require one photon to have an energy of at least 50 MeV
and the other to have an energy of at least 20 MeV.
We also require a narrow window around the di-photon invariant mass
to compensate for the lower photon-energy requirement:
within 10 MeV/$c^2$ of the nominal $\pi^0$ mass,
which corresponds to approximately $\pm 1.6 \sigma$.

%\subsection{D reconstruction}
Neutral $D$ mesons are reconstructed in the following decay modes:
$D^0 \rightarrow K^- \pi^+$,
$K_S^0 \pi^0$,
$K^+ K^-$,
$\pi^+ \pi^-$,
$K_S^0 \pi^+ \pi^-$,
$K^- \pi^+ \pi^0$,
$\pi^+ \pi^- \pi^0$,
$K_S^0 K^+ K^-$,
$K^- \pi^+ \pi^+ \pi^-$, and
$K_S^0 \pi^+ \pi^- \pi^0$.
Charged $D$ mesons are reconstructed in the following modes:
$D^+ \rightarrow K_S^0 \pi^+$,
$K^- \pi^+ \pi^+$,
$K_S^0 \pi^+ \pi^0$,
$K^+ K^- \pi^+$, and
$K_S^0 \pi^+ \pi^+ \pi^-$.
The combined reconstructed branching fractions are
37\% and 22\%
for $D^0$ and $D^+$, respectively.
For $D$ decay modes without a $\pi^0$ in the final state, 
we require the invariant mass of the $D$ candidates
to be within 15 MeV/$c^2$ of the $D^0$ or $D^+$ mass,
which corresponds to a window of approximately $\pm 3\sigma$.
For modes with a $\pi^0$ in the final state,
we require a wider invariant mass window:
from $-45$ to $+30$ MeV/$c^2$ around the nominal $D^0$ mass for $D^0$ candidates,
and 
from $-36$ to $+24$ MeV/$c^2$ around the nominal $D^+$ mass for $D^+$ candidates.
These windows correspond to approximately [$-1.2\sigma, +1.8\sigma$] and [$-1.0\sigma, +1.5\sigma$], respectively, in resolution.
Candidate $D^{*+}$ mesons are formed by combining $D^0$ and $\pi^+$ candidates
or $D^+$ and $\pi^0$ candidates.
%\subsection{D* reconstruction}
%The candidate $D^{*+}$ mesons are reconstructed in the $D^0 \pi^+$ and $D^+ \pi^0$.
To improve the resolution of the $D^*$-$D$ mass difference, $\Delta M$, for the $D^{*+} \rightarrow D^0 \pi^+$ decay mode,
the charged pion track from the $D^{*+}$ is refitted to the $D^0$ decay vertex.
We require $\Delta M$ to be within 2.5 MeV/$c^2$ and 2.0 MeV/$c^2$, respectively, around
the value of the nominal $D^*$-$D$ mass difference
for the $D^{*+} \rightarrow D^0 \pi^+$ and $D^{*+} \rightarrow D^+ \pi^0$ decay modes.
These windows correspond to $\pm 3.2 \sigma$ and $\pm 2.0 \sigma$, respectively, in resolution.
We apply a tighter window in the $D^{*+} \rightarrow D^+ \pi^0$ decay mode
to suppress a large contribution to the background arising from falsely reconstructed neutral pions.

\begin{figure}[htb]
  \includegraphics[width=7.8cm]{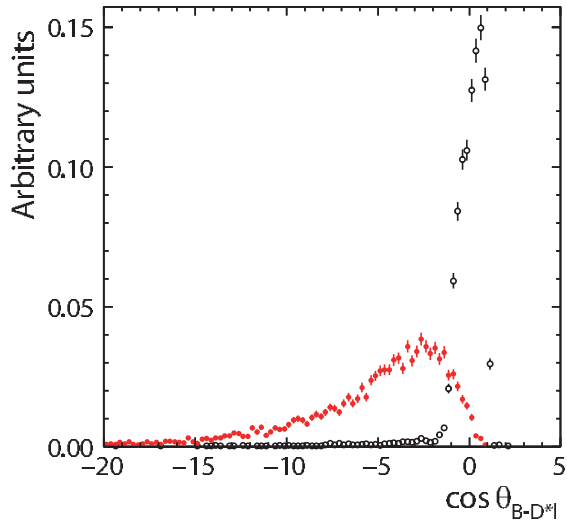}
  \caption{The $\cos \theta_{B \mathchar`-  D^* \ell}$ distributions for
  $\bar{B}^0 \rightarrow D^{*+} \tau^- \bar{\nu}_{\tau}$ (solid red circles) and
  $\bar{B}^0 \rightarrow D^{*+} \ell^- \bar{\nu}_{\ell}$ (open black circles)
  taken from MC simulation.}
  \label{fig:cos_bdstl}
\end{figure}

%\subsection{single semileptonic tag}
To tag semileptonic $B$ decays,
we combine $D^{*+}$ and lepton candidates
of opposite electric charge
and calculate the cosine of the angle between the momentum of the $B$ meson and the $D^* \ell$ system
in the $\Upsilon(4S)$ rest frame,
under the assumption that only one massless particle is not reconstructed:
\begin{eqnarray}
\cos \theta_{B \mathchar`-  D^* \ell} \equiv
\frac
{2E_{\rm beam} E_{D^* \ell} - m_B^2 c^4 - M_{D^* \ell}^2 c^4}
{2 |\vec{p}_B| \cdot |\vec{p}_{D^* \ell}| c^2},
\label{eq:cos_bdstrl}
\end{eqnarray}
where $E_{\rm beam}$ is the energy of the beam, and
$E_{D^* \ell}$, $\vec{p}_{D^* \ell}$, and $M_{D^* \ell}$
are the energy, momentum, and mass, respectively, of the $D^* \ell$ system.
The variable $m_B$ is the nominal $B$ meson mass~\cite{PDG}, and $\vec{p}_B$ is the nominal $B$ meson momentum.
All variables are defined in the $\Upsilon(4S)$ rest frame.
Figure~\ref{fig:cos_bdstl} shows the $\cos \theta_{B \mathchar`-  D^* \ell}$ distribution
for signal and normalization decay modes in MC samples.
Correctly reconstructed $B$ candidates in the normalization decay mode are expected to have
a value of $\cos \theta_{B \mathchar`-  D^* \ell}$ between $-1$ and $+1$.
Correctly reconstructed $B$ candidates in the signal decay mode
and falsely reconstructed $B$ candidates tend to have values of $\cos \theta_{B \mathchar`-  D^* \ell}$
below the physical region
due to contributions from additional particles.

%\subsection{double tag}
In each event we require two tagged $B$ candidates that are opposite in flavor.
Signal events may have the same flavor due to $B \bar{B}$ mixing;
however, we veto such events as they lead to ambiguous $D^* \ell$ pair assignment
and larger combinatorial background.
We require that at most one $B$ meson be reconstructed from a $D^+$ mode
to avoid large background from fake neutral pions when forming $D^*$ candidates.
In each signal event, we assign the candidate with the lower value of $\cos \theta_{B \mathchar`-  D^* \ell}$
(referred to hereinafter as $\cos \theta_{B \mathchar`-  D^* \ell}^{\rm sig}$)
as $B_{\rm sig}$.
The probability of falsely assigning the $B_{\rm sig}$ as the $B_{\rm tag}$ for signal events is about 3\%,
according to MC simulation.

After the identification of the $B_{\rm sig}$ and $B_{\rm tag}$ candidates,
we apply further background suppression criteria.
On the tag side,
we require $-2.0 < \cos \theta^{\rm tag}_{B \mathchar`-  D^* \ell} < +1.5$
to select $B \rightarrow D^* \ell \nu_{\ell}$.
On the signal side, we require the $D^*$ momentum in the $\Upsilon(4S)$ rest frame to be less than 2.0 GeV/$c$,
while, on the tag side,
we require it to be less than 2.5 GeV/$c$, which accounts for the lepton mass difference.
Finally, we require that events contain no extra charged tracks, $K_S^0$ candidates, or $\pi^0$ candidates,
which are reconstructed with the same criteria as those used for the $D$ candidates.
%\subsection{multiple candidate probability}

At this stage, the probability of finding multiple candidates is 7\%
which is mainly caused by swapped pions between signal and tag sides. 
When multiple candidates are found in an event,
we select a single candidate,
which has the smallest sum of two chi-square in vertex-constrained fits for the $D$ mesons,
among multiple candidates.
In the final sample,
the fraction of signal and normalization events are estimated to be 5\% and 68\%
from MC simulation.

\section{SIGNAL, NORMALIZATION AND BACKGROUND SEPARATION}
%\subsection{background rejection}
To separate reconstructed signal and normalization events,
we employ a neural network using the NeuroBayes software package
 \cite{NEUROBAYES}.
The variables used as inputs to the network are
$\cos \theta^{\rm sig}_{B \mathchar`-  D^* \ell}$,
the missing mass squared
$M_{\rm miss}^2 = (2E_{\rm beam} - \sum_i E_i)^2/c^4 - |\sum_i \vec{p}_i|^2/c^2$,
and
the visible energy $E_{\rm vis} = \sum_i E_i$,
where $(E_i, \vec{p}_i)$ is
the four-momentum of particle $i$ in the $\Upsilon(4S)$ rest frame.
The most powerful observable in separating signal and normalization is $\cos \theta^{\rm sig}_{B \mathchar`-  D^* \ell}$.
The neural network is trained using MC samples of signal and normalization events.
We will use the neural network classifier
as one of the fitting variables for the measurement of ${\cal R}(D^*)$
without any selection on the neural network classifier.
Typically, for a requirement the neural network classifier to be larger than 0.8,
82\% of the signal is kept while rejecting 97\% of the normalization events.

The dominant background contributions arise from 
events with misreconstructed $D^{(*)}$ mesons (denoted fakes).
The sub-dominant contributions arise from two sources
in which $D^*$ mesons from both $B_{\rm sig}$ and $B_{\rm tag}$ are correctly reconstructed.
One source is $B \rightarrow D^{**} \ell \nu_{\ell}$,
where the $D^{**}$ meson decays to $D^{(*)}$ and other particles.
%, which is one pion in many case of 85\% or more.
The other source is $B \rightarrow X_c D^*$ events,
where one $D^*$ meson is correctly reconstructed
and the other charmed meson $X_c$ decays semileptonically.
If the hadrons in the semileptonic $X_c$ decay are not identified, such events can mimic signal.
Similarly, events in which $X_c$ is a $D_s^+$ meson decaying into $\tau^+ \nu_{\tau}$ can also mimic signal.

\begin{figure}[htb]
  \includegraphics[width=7.8cm]{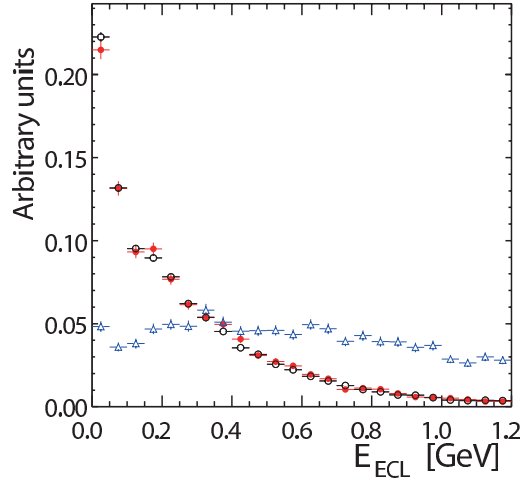}
  \caption{The $E_{\rm ECL}$ distributions for the signal (solid red circles),
  the normalization (open black circles), and the background (open blue triangles)
  taken from MC simulation,
  where the $E_{\rm ECL}$ is defined as the sum of the energies of neutral clusters
  detected in the ECL that are not associated with reconstructed particles.
  }
  \label{fig:mc_eecl_dist}
\end{figure}

To separate signal and normalization events from background processes,
we place a criterion on the sum of the energies of neutral clusters
detected in the ECL that are not associated with reconstructed particles,
denoted as $E_{\rm ECL}$.
To mitigate the effects of photons related to beam background in the energy sum,
we only include clusters with energies greater than 50, 100, and 150 MeV, respectively,
from the barrel, forward, and backward calorimeter regions, defined in Ref.~\cite{BELLE}.
Signal and normalization events peak near zero in $E_{\rm ECL}$,
while background events can populate a wider range
as shown in Figure~\ref{fig:mc_eecl_dist}.
We require $E_{\rm ECL}$ to be less than 1.2 GeV.
%This selection criterion is optimized to maximize the sensitivity in a signal region
%${\cal O}_{\mathit{NB}} > 0.8$ and $E_{\rm ECL} < 0.5$ GeV.

\section{MC CALIBRATION}
To improve the accuracy of the MC simulation,
we apply a series of calibration factors determined from control sample measurements.
The lepton identification efficiencies are separately corrected for electrons and for muons 
to account for differences between the detector responses in data and MC.
Correction factors for lepton identification efficiencies are evaluated
as a functions of the momentum and direction of the lepton
using $e^+ e^- \rightarrow e^+ e^- \ell^+ \ell^-$ and $J/\psi \rightarrow \ell^+ \ell^-$ decays.
We reweight events to account for differing $D^{(*)}$ yields between data and MC.

The differing yields of correctly reconstructed $D^{(*)}$ mesons
in data and MC affect the ${\cal R}(D^*)$ measurement,
as it biases the determination of the background contribution.
Calibration factors for events with both correctly- and falsely-reconstructed $D$ mesons
are estimated for each $D$ meson decay mode
using a two-dimensional fit to $M_D$.
For this calibration,
we use samples with all selection criteria other than
$D$ mass and $\Delta M$ applied.
%Precise calibration can be performed
%by using samples with two tagged candidates,
%which have good purity and are close to final samples for the ${\cal R}(D^*)$ measurement.
A two-dimensional PDF is constructed
by taking the product of the one-dimensional functions
%for $M_D$(tag) and $M_D$(sig) respectively.
for $M_D$.
The PDF in each dimension is the sum of a signal component
%modeled by a triple Gaussian or Crystal Ball function \cite{CB_FUNCTION} plus a Gaussian
and a background component modelled with a first-order polynomial.
The signal component is a triple Gaussian for $D^0$ decay modes without a $\pi^0$
and
a Crystal Ball function \cite{CB_FUNCTION} plus a Gaussian for $D^0$ decay modes with a $\pi^0$ and $D^+$ decay modes.
In this calibration, we do not distinguish signal and tag sides.
To estimate calibration factors for specific $D$ decay modes,
we fit samples in which
one $D$ meson is reconstructed in a specific mode
while the other is reconstructed in any signal mode.
%From the signal and background yield ratios of data to MC samples,
From the ratios of data to MC samples in signal and background yields,
we derive calibration factors of the specific decay mode for events with correctly and falsely reconstructed $D$ mesons.
We cannot independently determine calibration factors for all $D$ meson decay modes
as we use other decay modes
when we calibrate each specific decay mode of a given $D$ meson.
To estimate all the calibration factors correctly,
we first perform the two-dimensional fit for each decay mode without weighting factors,
and then iterate the fits using resultant weighting factors
until all calibration factors converge.

Similarly, we estimate calibration factors for events with correctly and falsely reconstructed $D^*$ mesons
from a two-dimensional fit to $\Delta M$.
%(tag, and sig).
Calibration factors for events with correctly and falsely reconstructed $D^*$ mesons
are separately estimated for subsequent decay to $D^0$ and $D^+$ mesons.
For this calibration, we use samples in which one $D^*$ meson is reconstructed from $D^0 \pi^+$
and the other $D^*$ meson is reconstructed from $D^+ \pi^0$.
We apply derived calibration factors to samples, in which
both $D^*$ mesons are reconstructed from $D^0 \pi^+$,
and find good agreement.
Eventually, the deviations of the yields between the MC sample and the data
reduce
from 1.1$\sigma$ to 0.2$\sigma$ for the yields of correctly reconstructed $D^*$ mesons and
from 8.7$\sigma$ to 0.3$\sigma$ for the yields of falsely   reconstructed $D^*$ mesons,
where $\sigma$ is quadratic sum of statistical error from two-dimensional fit to $\Delta M$ in data and MC. 

\section{MAXIMUM LIKELIHOOD FIT}
We extract the yields of the signal and normalization processes
from a two-dimensional extended maximum-likelihood fit to
neural network classifier output ${\cal O}_{\mathit{NB}}$ and $E_{\rm ECL}$.
The likelihood function consists of five components:
signal, normalization, fake $D^{(*)}$ events, $B \rightarrow D^{**} \ell \nu_{\ell}$,
and other backgrounds (predominantly from $B \rightarrow X_c D^*$).
The PDFs of all components are determined from MC simulation.
There are significant correlations between ${\cal O}_{\mathit{NB}}$ and $E_{\rm ECL}$
in the normalization and background components, but not for the signal. 
We therefore construct the normalization and background PDFs using two-dimensional histograms
and apply a smoothing procedure to account for its limited statistical power \cite{SMOOTHING}.
The signal PDF is the product of one-dimensional histograms
in ${\cal O}_{\mathit{NB}}$ and $E_{\rm ECL}$.

%%%%%%%%%%%%%
% FIT PARAM %
%%%%%%%%%%%%%
Three parameters are floated in the final fit:
the yields of the signal, normalization, and $B \rightarrow D^{**} \ell \nu_{\ell}$ components.
The yield of fake $D^{(*)}$ events is fixed
to the value
estimated from sidebands in the $\Delta M$ distribution.
Since the PDF shape of fake $D^{(*)}$ events depends on the composition of
signal, normalization, $B \rightarrow D^{**} \ell \nu_{\ell}$, and other backgrounds,
the relative contributions of these processes to the fake $D^{(*)}$ component are described as a function of the three fit parameters. 
The yields of other backgrounds
are fixed to the values expected from MC simulation.
The ratio ${\cal R}(D^*)$ is given by the formula:
\begin{eqnarray}
{\cal R}(D^*) &=&
\frac{1}
{
2{\cal B}(\tau^- \rightarrow \ell^- \bar{\nu}_{\ell} \nu_{\tau})
}
\cdot
\frac{\varepsilon_{\rm norm}}{\varepsilon_{\rm sig}}
\cdot
\frac{N_{\rm sig}}{N_{\rm norm}},
\label{eq:cal_rdstr}
\end{eqnarray}
where
$\varepsilon_{\rm sig (norm)}$ and $N_{\rm sig (norm)}$
are the reconstruction efficiency and yields
of signal (normalization) events.
We use ${\cal B}(\tau^- \rightarrow \ell^- \bar{\nu}_{\ell} \nu_{\tau}) = 0.176 \pm 0.003$
as the average of the world averages for $\ell =e$ and $\ell = \mu$~\cite{PDG}.
The ratio of efficiencies, $\varepsilon_{\rm norm}/\varepsilon_{\rm sig}$, is estimated to be $1.289 \pm 0.015$ from MC simulation.
The difference between reconstruction efficiencies of signal and normalization events arises from their distinct lepton momentum distributions,
and the different event criteria on the $D^*$ momenta.

To validate the fit procedure,
we perform the fitting to multiple subsets of the available MC samples.
Furthermore, we validate the fit procedure
by a large number of pseudo experiments.
We have not observed any bias.

\section{PDF VALIDATION}
We validate the PDFs used in the fitting procedure by analysing various control samples.
For fake $D^{(*)}$ events, we study the $\Delta M$ sidebands,
where we find good agreement in both ${\cal O}_{\mathit{NB}}$ and $E_{\rm ECL}$.
For $B \rightarrow D^* \ell \nu_{\ell}$ decays, we require one $B$ meson to be reconstructed with the hadronic tagging method~\cite{HADTAG_NEURO}
and the other $B$ meson  reconstructed with the nominal criteria of this analysis.
We find good agreement between data and MC in the
$E_{\rm ECL}$, $M_{\rm miss}^2$, and $E_{\rm vis}$ distributions,
but small discrepancies in the $\cos \theta_{B \mathchar`-  D^* \ell}$ distributions~\cite{SUPP},
which we incorporate as a systematic uncertainty.

\section{SYSTEMATIC UNCERTAINTIES}
To estimate the systematic uncertainties on ${\cal R}(D^*)$,
we vary every fixed parameter in turn by one standard deviation
and repeat the fit.
The systematic uncertainties are summarized in Table~\ref{tab:systematics}.
The dominant systematic uncertainty arises from
the limited size of the MC samples: to estimate this uncertainty,
we recalculate PDFs for signal, normalization, fake $D^{(*)}$ events, $B \rightarrow D^{**} \ell \nu_{\ell}$, and other backgrounds
by generating toy MC samples from the nominal PDFs according to Poisson statistics
and repeat the fit with the new PDFs.

Small discrepancies between the data and MC are found in the $\cos \theta_{B \mathchar`-  D^* \ell}$ distributions in the hadronic tagged samples.
We estimate it as ``PDF shape of the normalization in $\cos \theta_{B \mathchar`-  D^* \ell}$''
in Table~\ref{tab:systematics}
by correcting the $\cos \theta_{B \mathchar`-  D^* \ell}$ distribution in MC samples
according to the observed discrepancies,
and repeating the fit.

The branching fractions of the $B \rightarrow D^{**} \ell \nu_{\ell}$ decay modes and the decays of the $D^{**}$ mesons are not well known
and therefore contribute significantly
to the total PDF uncertainty for $B \rightarrow D^{**} \ell \nu_{\ell}$ decays.
The branching fraction of each $B \rightarrow D^{**} \ell \nu_{\ell}$ decay
is varied within its uncertainty.
The uncertainties are assumed
to be
$\pm 6\%$ for $D_1$, 
$\pm 10\%$ for $D_2^*$,
$\pm 83\%$ for $D_1'$,
and
$\pm 100\%$ for $D_0^*$,
including the limited knowledge of the $D^{**}$ decays.
We also consider the impact of contributions from radially excited $D(2S)$ and $D^*(2S)$,
where we assume the branching fractions of $B \rightarrow D^{(*)}(2S) \ell \nu_{\ell}$
to be as large as $0.5\%$. 

The yield of fake $D^{*}$ events is fixed
to the value estimated from sidebands in the $\Delta M$ distribution.
We vary this yield within its uncertainties.
We also vary the calibration factors for $D$ meson decay modes
within their uncertainties for events with
falsely reconstructed $D^{(*)}$ events.
%in order to take into account possible dependence of PDF shape to $D$ meson decay mode.

The yields of other background processes, predominantly from $B \rightarrow X_c D^*$ events, are fixed
to the values estimated from MC simulation.
We consider variations on the yield and shape of the PDF of these background processes
within their  measured uncertainties.
The uncertainties for the $B \rightarrow X_c D^*$ channels are assumed
to be
$\pm  8\%$ for $B \rightarrow D_s^*  D^{*-}$,
$\pm 14\%$ for $B \rightarrow D_s    D^{*-}$,
$\pm  8\%$ for $B \rightarrow D^{*+} D^{*-}$, and
$\pm 10\%$ for $B \rightarrow D^+    D^{*-}$.
Furthermore,
we add an uncertainty of $\pm 4\%$ due to the size of the MC sample.
We determine the uncertainty from the branching fraction of $D_s \rightarrow \tau \nu_{\tau}$ decay
(which may peak near the signal in the $E_{\rm ECL}$ distribution)
to be negligible.

The reconstruction efficiency ratio of signal to normalization events
is varied within its uncertainty,
which is limited by the size of the MC samples for signal events.

We include other minor systematic uncertainties from two sources:
one related to the parameters
that are used for the reweighting of the semileptonic $B \rightarrow D^{(*(*))} \ell \nu_{\ell}$ decays
from the ISGW model to the LLSW model;
and the other from the branching fraction of
$\tau^- \rightarrow \ell^- \bar{\nu}_{\ell} \nu_{\tau}$ decay \cite{PDG}.
The total systematic uncertainty is estimated by summing the above uncertainties in quadrature.

\begin{table*}[htbp]
\caption{Summary of the systematic uncertainties on ${\cal R}(D^*)$ for electron and muon modes combined and separated.
The uncertainties are relative and are given in percent.}
\begin{center}
\begin{tabular}{c|c|c|c} \hline
& \multicolumn{3}{c}{${\cal R}(D^*)$ [\%]} \\ \hline
Sources & $\ell^{\rm sig} = e,\mu$ & $\ell^{\rm sig} = e$ & $\ell^{\rm sig} = \mu$ \\ \hline
MC size for each PDF shape                                                &  2.2 & 2.5  & 3.9 \\[1.0pt]
PDF shape of the normalization in $\cos \theta_{B \mathchar`-  D^* \ell}$       &  $^{+1.1}_{-0.0}$ & $^{+2.1}_{-0.0}$  & $^{+2.8}_{-0.0}$\\[3.0pt]
PDF shape of $B \rightarrow D^{**} \ell \nu_{\ell}$                             &  $^{+1.0}_{-1.7}$ & $^{+0.7}_{-1.3}$ & $^{+2.2}_{-3.3}$ \\[1.0pt]
PDF shape and yields of fake $D^{(*)}$                                          &  1.4 & 1.6  & 1.6 \\
PDF shape and yields of $B \rightarrow X_c D^*$                                 &  1.1 & 1.2  & 1.1 \\
Reconstruction efficiency ratio $\varepsilon_{\rm norm}/\varepsilon_{\rm sig}$  &  1.2 & 1.5  & 1.9 \\ 
Modeling of semileptonic decay                                                  &  0.2 & 0.2  & 0.3 \\
${\cal B}(\tau^- \rightarrow \ell^- \bar{\nu}_{\ell} \nu_{\tau})$               &  0.2 & 0.2  & 0.2 \\ \hline
Total systematic uncertainty                                                  &  $^{+3.4}_{-3.5}$ & $^{+4.1}_{-3.7}$ & $^{+5.9}_{-5.8}$ \\ \hline
%Statistical uncertainty                                                       & 10.1 & 12.3 & 16.9 \\ \hline
\end{tabular}
\label{tab:systematics}
\end{center}
\end{table*}

\section{RESULTS}
\begin{figure*}[htb]
\centering
\subfigure{
\includegraphics*[width=5.6cm]{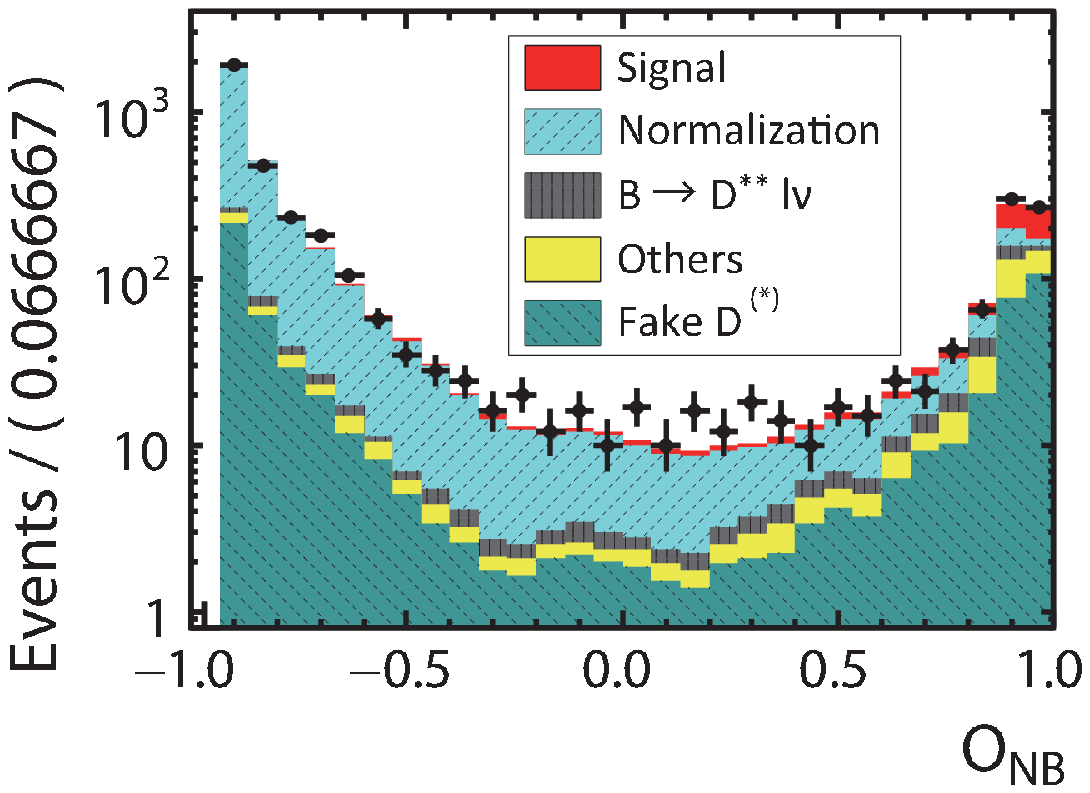}
\label{fig:result_lep0_nb}}
\subfigure{
\includegraphics*[width=5.6cm]{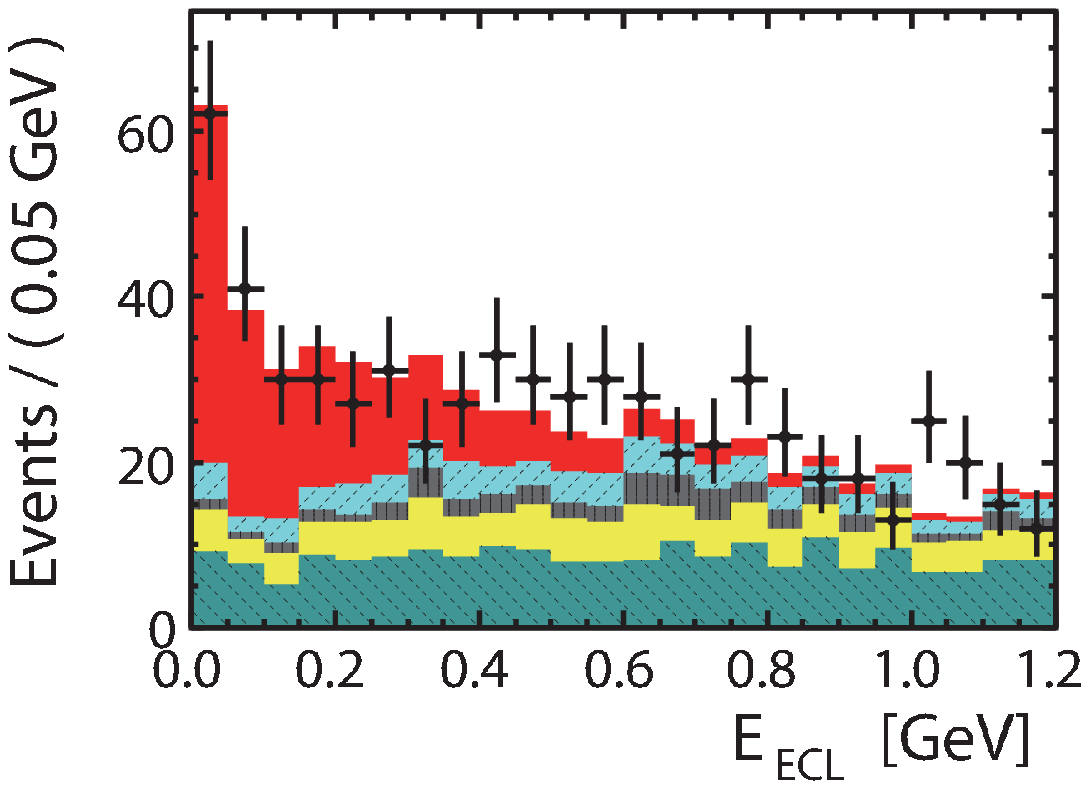}
\label{fig:result_lep0_eecl_sigenh}}
\subfigure{
\includegraphics*[width=5.6cm]{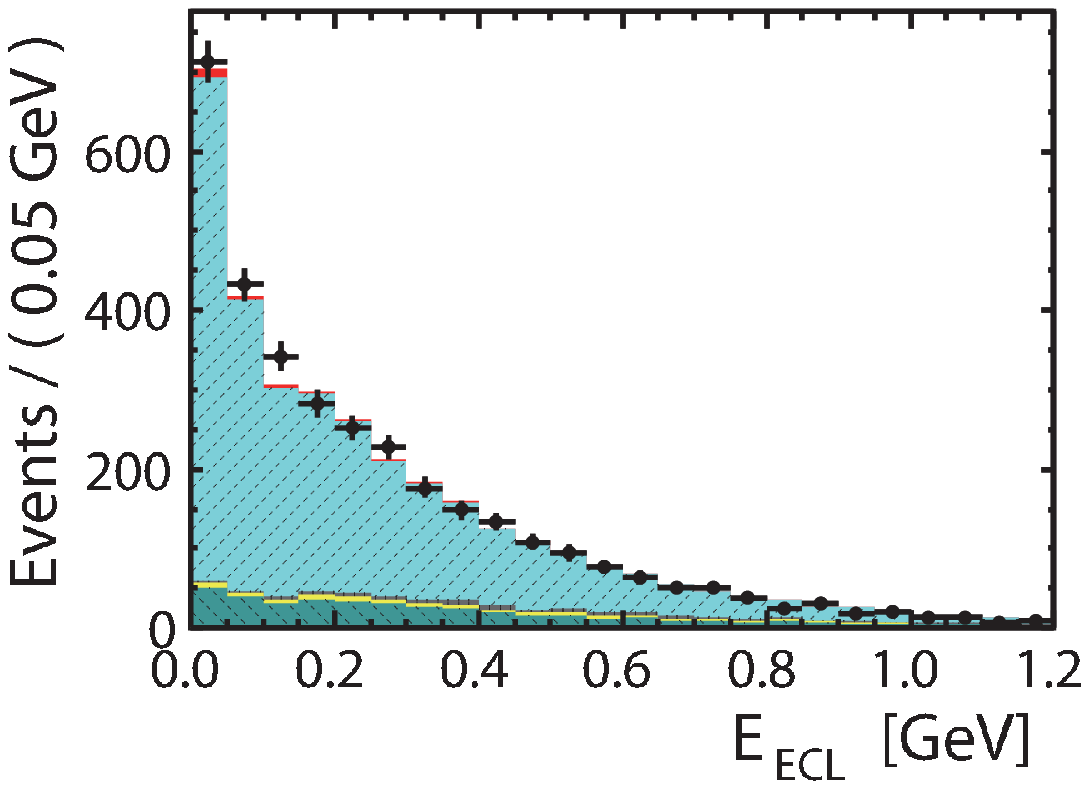}
\label{fig:result_lep0_eecl_normenh}}
\caption{Projections of the fit results with data points overlaid for
(left) the neural network classifier output, ${\cal O}_{\mathit{NB}}$, and
the $E_{\rm ECL}$ distribution in (center) the signal-enhanced region, ${\cal O}_{\mathit{NB}} > 0.8$, and
(right) the normalization-enhanced region, ${\cal O}_{\mathit{NB}} < 0.8$.
The background categories are described in detail in the text,
where ``others'' refers to predominantly $B \rightarrow X_c D^*$ decays.
}
\label{fig:result_fit}
\end{figure*}

The ${\cal O}_{\mathit{NB}}$ and $E_{\rm ECL}$ projections of the fitted distributions are shown in Figure~\ref{fig:result_fit}.
The yields of signal and normalization events are measured to be $231 \pm 23({\rm stat})$ and $2800 \pm 57({\rm stat})$, respectively.
The ratio
${\cal R}(D^*)$ is found to be
\begin{eqnarray}
{\cal R}(D^*) &=& 0.302 \pm 0.030 \pm 0.011,
\end{eqnarray}
where
the first uncertainty is statistical and
the second systematic (and likewise for all following results).

We calculate the statistical significance of the signal as $\sqrt{-2\ln ({\cal L}_0/{\cal L}_{\rm max})}$,
where ${\cal L}_{\rm max}$ and ${\cal L}_0$
are the maximum likelihood and the likelihood obtained
assuming zero signal yield, respectively.
We obtain a statistical significance of $13.8\sigma$.
We also estimate the compatibility of the measured value of ${\cal R}(D^*)$
and the SM prediction.
%by fixing the value of ${\cal R}(D^*)$ to the SM prediction.
The effect of systematic uncertainties is included
by convolving the likelihood function
with a Gaussian distribution.
%whose width corresponds to the systematic uncertainty of ${\cal R}(D^*)$.
We conclude that our result is larger than the SM prediction
by $1.6\sigma$.

\section{CROSS-CHECKS}
To determine the consistency of the measured value of ${\cal R}(D^*)$ among $\tau$ final states, 
we divide the data samples by lepton flavor on the signal side
and fit them separately.
All PDFs for electron and muon channels are separately constructed from the MC samples.
The efficiency ratios $\varepsilon_{\rm norm}/\varepsilon_{\rm sig}$
are estimated to be
$1.107 \pm 0.016$ and $1.591 \pm 0.030$
for electron and muon channels of the tau decays, respectively.
We obtain
\begin{eqnarray}
{\cal R}(D^*) &=& 0.311 \pm 0.038 \pm 0.013 \hspace{0.5em}(\ell^{\rm sig}=e), \\
{\cal R}(D^*) &=& 0.304 \pm 0.051 \pm 0.018 \hspace{0.5em}(\ell^{\rm sig}=\mu).
\end{eqnarray}
The systematic uncertainties are summarized in Table~\ref{tab:systematics}.
These two results are consistent with each other.

To study $B \rightarrow D^{**} \ell \nu_{\ell}$ background contributions,
we require an additional $\pi^0$ with respect to the nominal event selection.
In this control sample,
we calculate $E'_{\rm ECL}$,
which is defined as the remaining energy
after the energy deposit from the additional $\pi^0$ is removed from $E_{\rm ECL}$.
The $B \rightarrow D^{**} \ell \nu_{\ell}$ background contributions are extracted
from the control samples using the nominal fitting method,
replacing $E_{\rm ECL}$ with $E'_{\rm ECL}$,
which is defined as $E_{\rm ECL}$ without the energy deposit from the additional $\pi^0$~\cite{SUPP}.
We find consistent results for the branching fractions of $B \rightarrow D^{**} \ell \nu_{\ell}$
in the control and signal samples.

\section{NEW PHYSICS COMPATIBILITY TESTS}
Assuming all neutrinos are left-handed, %(or anti-right handed),
the effective Hamiltonian that contains
all possible four-fermion operators for the $b \rightarrow c \tau \nu_{\tau}$ decay
can be described as follows~\cite{SIG_DECAY_MODEL}:
\begin{eqnarray}
{\cal H}_{\rm eff} &=& \frac{4G_F}{\sqrt{2}}V_{cb}
\left[
{\cal O}_{V_1} +
\sum_{X = S_1, S_2, V_1, V_2, T}
C_X {\cal O}_X
\right],
\end{eqnarray}
where the four-Fermi operators, ${\cal O}_X$, are defined as
\begin{eqnarray}
{\cal O}_{S_1} &=& (\bar{c}_L                  b_R) (\bar{\tau}_R                  \nu_{\tau L}), \\
{\cal O}_{S_2} &=& (\bar{c}_R                  b_L) (\bar{\tau}_R                  \nu_{\tau L}), \\
{\cal O}_{V_1} &=& (\bar{c}_L \gamma^{\mu}     b_L) (\bar{\tau}_L \gamma_{\mu}     \nu_{\tau L}), \\
{\cal O}_{V_2} &=& (\bar{c}_R \gamma^{\mu}     b_R) (\bar{\tau}_L \gamma_{\mu}     \nu_{\tau L}), \\
{\cal O}_{T  } &=& (\bar{c}_R \sigma^{\mu \nu} b_L) (\bar{\tau}_R \sigma_{\mu \nu} \nu_{\tau L}),
\end{eqnarray}
and  the $C_X$ parameters are the Wilson coefficients of ${\cal O}_X$.
We investigate the compatibility of the data samples
with new physics using a model-independent approach,
separately examining the impact of each operator.
In each new-physics scenario, we take into account changes
in the efficiency and fit PDF shapes using dedicated signal simulation.
We set the Wilson coefficients to be real in all cases.
Since ${\cal O}_{V_1}$ is just the SM operator,
it would change only ${\cal R}(D^{*})$, but not the kinematic distributions.
In the type-II two-Higgs doublet model (2HDM),
the relevant Wilson coefficients are given as
$C_{S_1} = -m_b m_\tau \tan^2 \beta/m_{H^+}^2$ and
$C_{S_2} = -m_c m_\tau /m_{H^+}^2$,
where $\tan \beta$ is the ratio of the vacuum expectation values of the two Higgs doublets,
and $m_b$, $m_c$, $m_{\tau}$, and $m_{H^+}$ are the masses of the $b$ quark, $c$ quark, $\tau$ lepton, and charged Higgs boson.
Since the contribution from $C_{S_2}$ is almost negligibly small except for the light charged Higgs boson,
we neglect the contribution from $C_{S_2}$ in the type-II 2HDM.

Various leptoquark models have been presented to explain anomalies in ${\cal R}(D^{(*)})$ in Ref.~\cite{LQ1}.
In addition to the model-independent study,
we study two representative models: $R_2$ and $S_1$.
Model $R_2$ contains scalar leptoquarks
of the type $(3,2)_{7/6}$ using the notation $(SU(3)_c, SU(2)_L)_Y$,
where $SU(3)_c$ is the representation under the generators of QCD,
$SU(2)_L$ is the representation under the generators of weak isospin,
and $Y$ is the weak hypercharge.
Model $S_1$ contains leptoquarks of the type $(3^*,1)_{1/3}$.
In these leptoquark models,
the relevant Wilson coefficients are related by 
$C_{S_2} = + 7.8 C_T$ for the $R_2$-type leptoquark model
and
$C_{S_2} = - 7.8 C_T$ for the $S_1$-type leptoquark model
at the $b$ quark mass scale,
assuming a leptoquark mass scale of 1 TeV/$c^2$.
Although the $V_1$ operator can appear independently of the $S_2$ and $T$ operators in the $S_1$-type leptoquark model,
we assume no contribution from the $V_1$ operator in this study.

\begin{figure*}[htb]
\centering
\subfigure{
\includegraphics*[width=7.5cm]{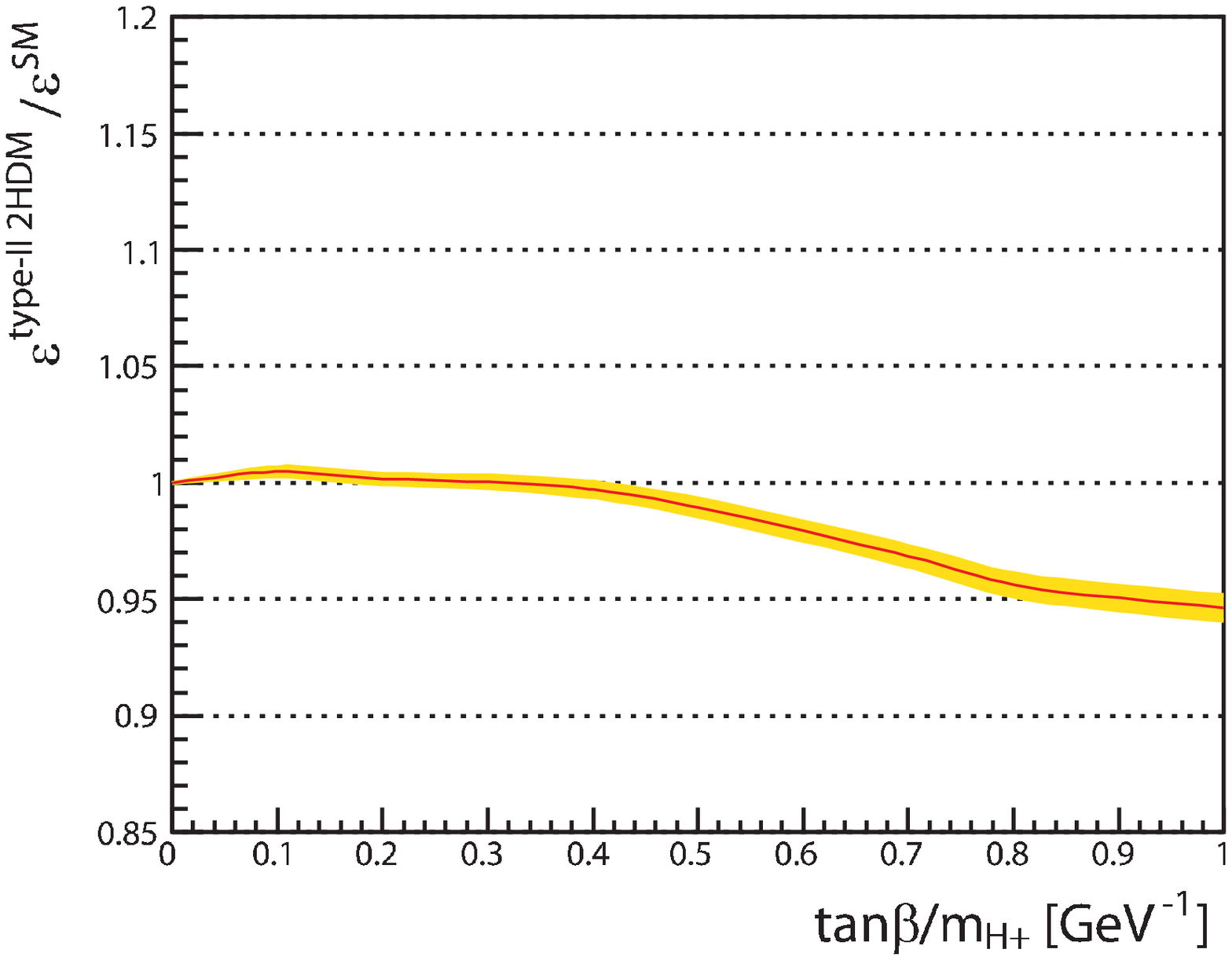}
\label{fig:npcurve_eff_2hdmII}}
\subfigure{
\includegraphics*[width=7.5cm]{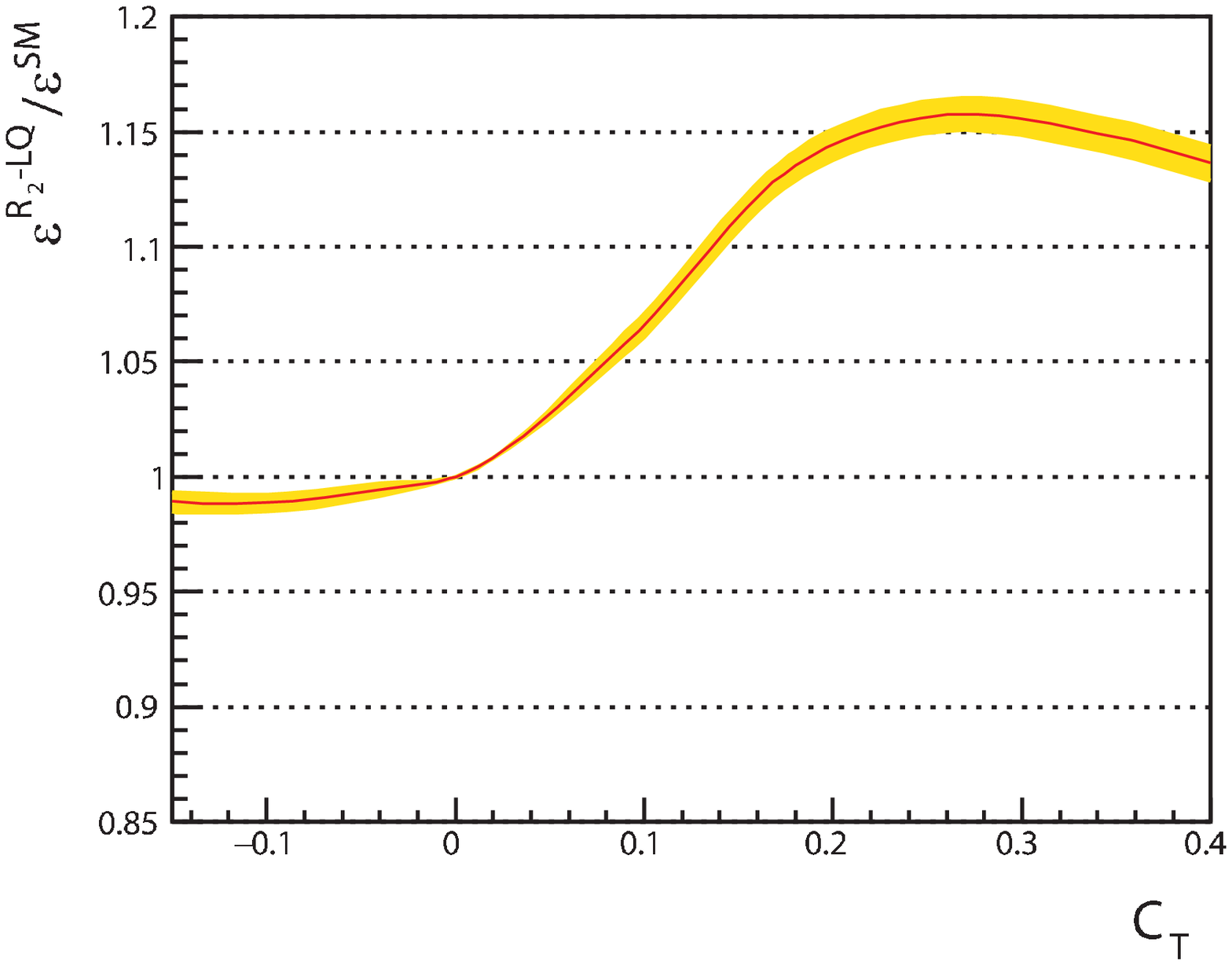}
\label{fig:npcurve_eff_R2LQ}}
\caption{The efficiencies for (left) the type-II 2HDM and (right) $R_2$-type leptoquark model with respect to the SM value.}
\label{fig:npcurve_eff}
\end{figure*}

\begin{figure*}[htb]
\centering
\subfigure{
\includegraphics*[width=7.5cm]{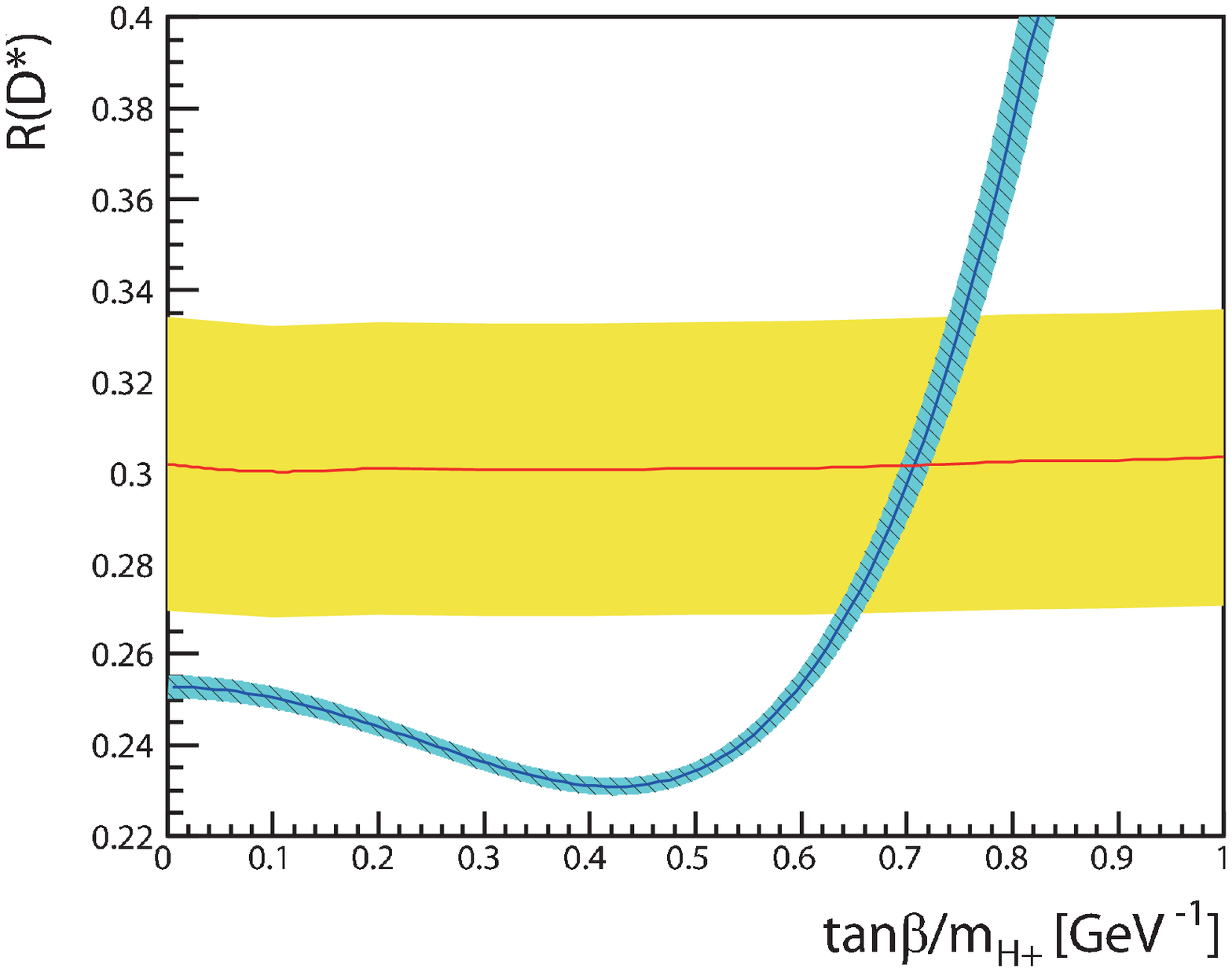}
\label{fig:npcurve_rdstr_2hdmII}}
\subfigure{
\includegraphics*[width=7.5cm]{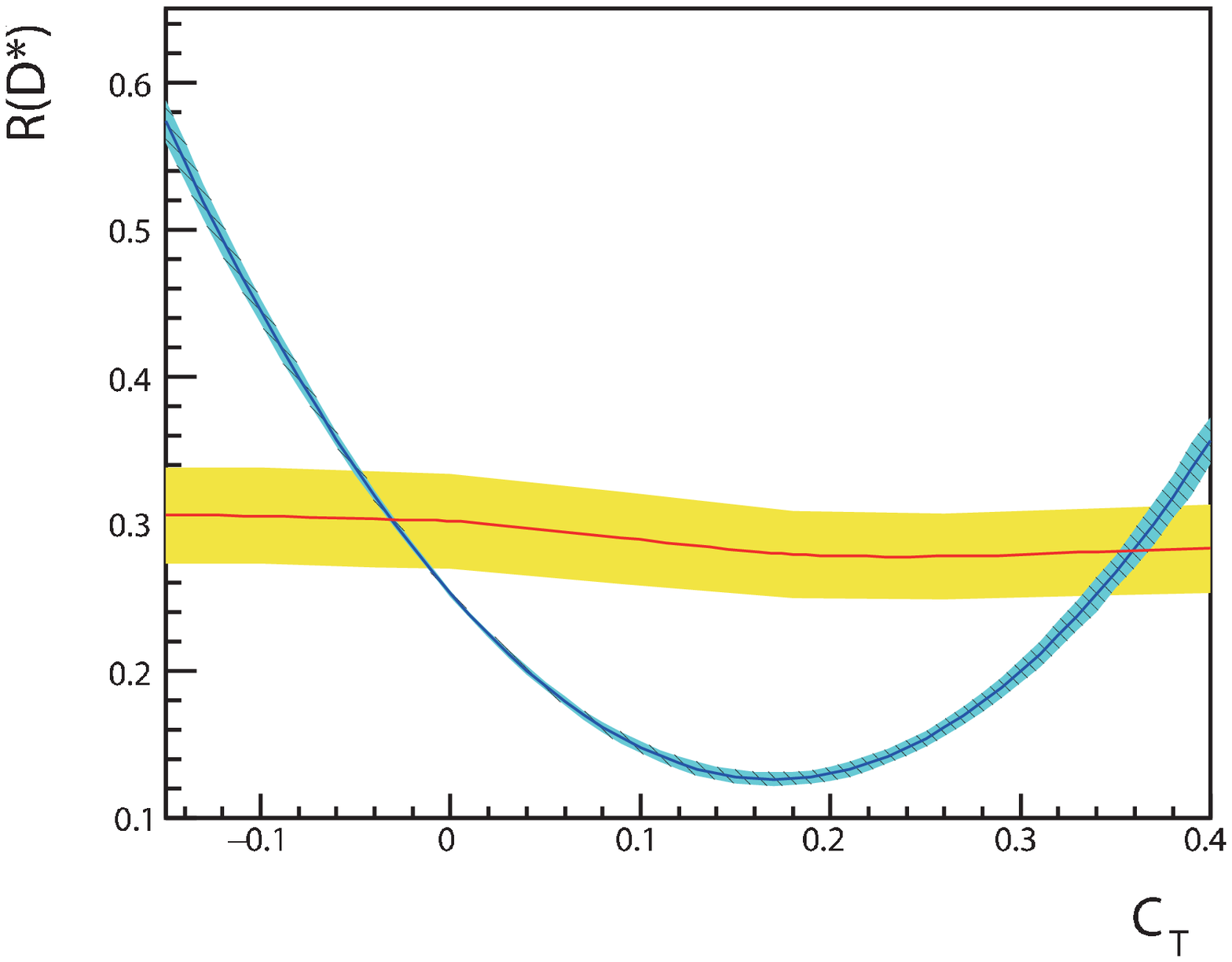}
\label{fig:npcurve_rdstr_R2LQ}}
  \caption{The measured values of ${\cal R}(D^*)$ for (left) the type-II 2HDM and (right) $R_2$-type leptoquark models,
  where central values are given as the solid (red) curves
  and the $1\sigma$ uncertainties are given as the shaded (yellow) regions.
  The theoretical predictions and their $1\sigma$ uncertainties are shown
as solid (blue) curves and hatched (light blue) regions, respectively~\cite{SIG_DECAY_MODEL}.}
  \label{fig:npcurve_rdstr}
\end{figure*}

Figure~\ref{fig:npcurve_eff} shows the dependence of the efficiency and measured value of ${\cal R}(D^*)$
as a function of the values of the respective parameters in the type-II 2HDM and the $R_2$-type leptoquark model.
Efficiency variations for other scenarios are shown in Ref.~\cite{SUPP}.
We find that efficiencies increase by up to 17\% for ${\cal O}_{V_2}$ and ${\cal O}_T$,
mainly due to the variation of the $D^*$ momentum distribution.
Similarly,
the efficiencies increase by up to 16\% and 11\%
in $R_2$- and $S_1$-type leptoquark models, respectively,
which include contributions from ${\cal O}_T$.
In other scenarios, the efficiency variation is 6\% or less.
Figure~\ref{fig:npcurve_rdstr}
shows the dependency of the measured values of ${\cal R}(D^*)$
on the values of the respective parameters in the type-II 2HDM and the $R_2$-type leptoquark model.
The allowed regions with 68\% confidence level (C.L.) of the respective parameters are summarized in Table~\ref{tab:np_favored_region}.

\begin{table*}[htbp]
\caption{Allowed regions with 68\% C.L. of Wilson coefficients~\cite{SIG_DECAY_MODEL}.
$-4.25 < C_{S_1} < -3.09$ corresponds to
$0.65$ GeV$^{-1} < \tan \beta / m_{H^+} < 0.76$ GeV$^{-1}$ in type-II 2HDM,
where $m_b = 4.20$ GeV/$c^2$, $m_c = 0.901$ GeV/$c^2$~\cite{QUARK_MASS} and $m_{\tau} = 1.77682$ GeV/$c^2$ \cite{PDG} are used.
}
\begin{center}
\begin{tabular}{c|c|c|c} \hline \hline
Models or operators    & Parameters                       & Allowed regions \\
                       &                                  & (68\% C.L.)    \\ \hline

${\cal O}_{S_1}$       & $C_{S_1}$                        & $[-4.25, -3.09], [+0.44, +1.57]$ \\
${\cal O}_{S_2}$       & $C_{S_2}$                        & $[-1.56, -0.43], [+3.12, +4.28]$ \\
${\cal O}_{V_1}$       & $C_{V_1}$                        & $[-2.15, -2.03], [+0.05, +0.15]$ \\
${\cal O}_{V_2}$       & $C_{V_2}$                        & $[-0.17,  0.00], [+1.83, +1.96]$ \\
${\cal O}_{T}$         & $C_{T}$                          & $[-0.06, -0.01], [+0.34, +0.39]$ \\
$R_2$-type leptoquark  & $C_T (= +C_{S_2}/7.8)$           & $[-0.05, -0.01], [+0.34, +0.38]$ \\
$S_1$-type leptoquark  & $C_T (= -C_{S_2}/7.8)$           & $[-0.07, -0.01], [+0.22, +0.28]$ \\
\hline \hline
\end{tabular}
\label{tab:np_favored_region}
\end{center}
\end{table*}

\begin{figure*}[htb]
\centering
\subfigure{
\includegraphics*[width=5cm]{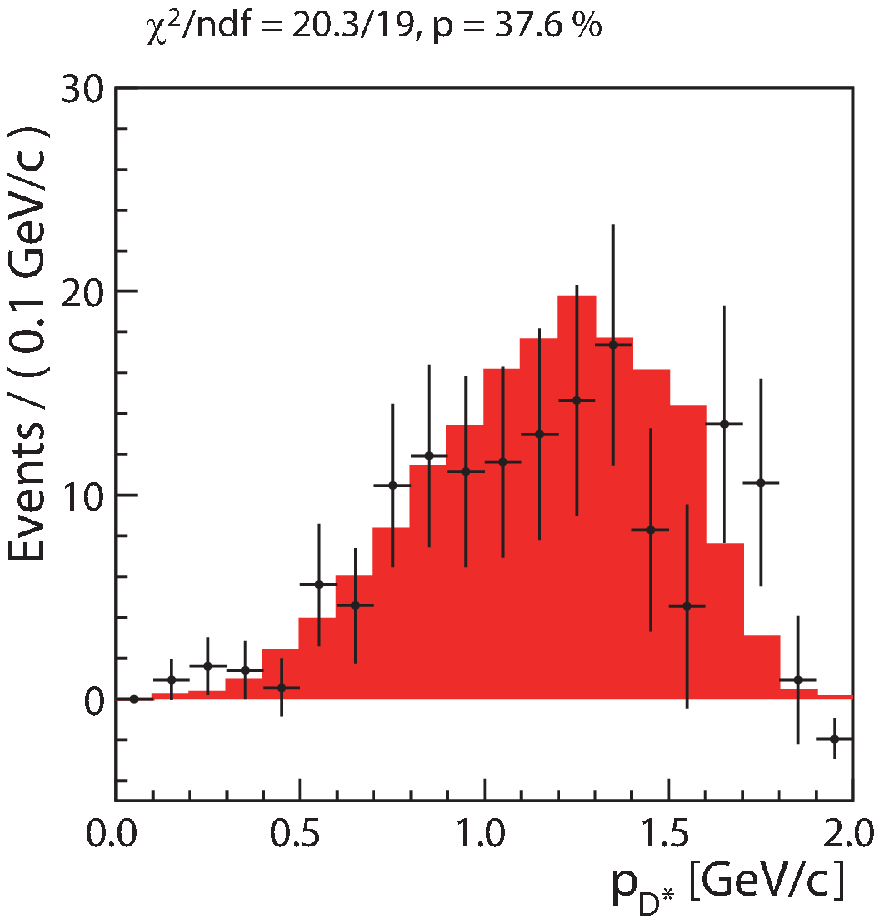}
\label{fig:kinematics_dstr_SM}}
\subfigure{
\includegraphics*[width=5cm]{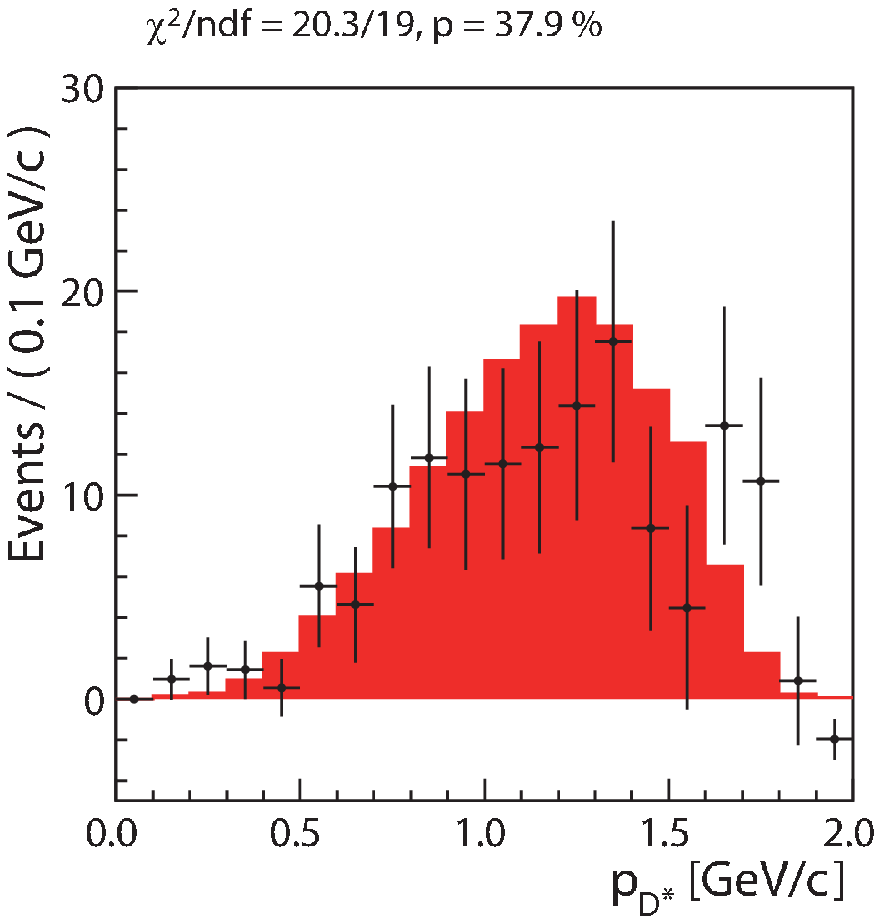}
\label{fig:kinematics_dstr_2hdmII}}
\subfigure{
\includegraphics*[width=5cm]{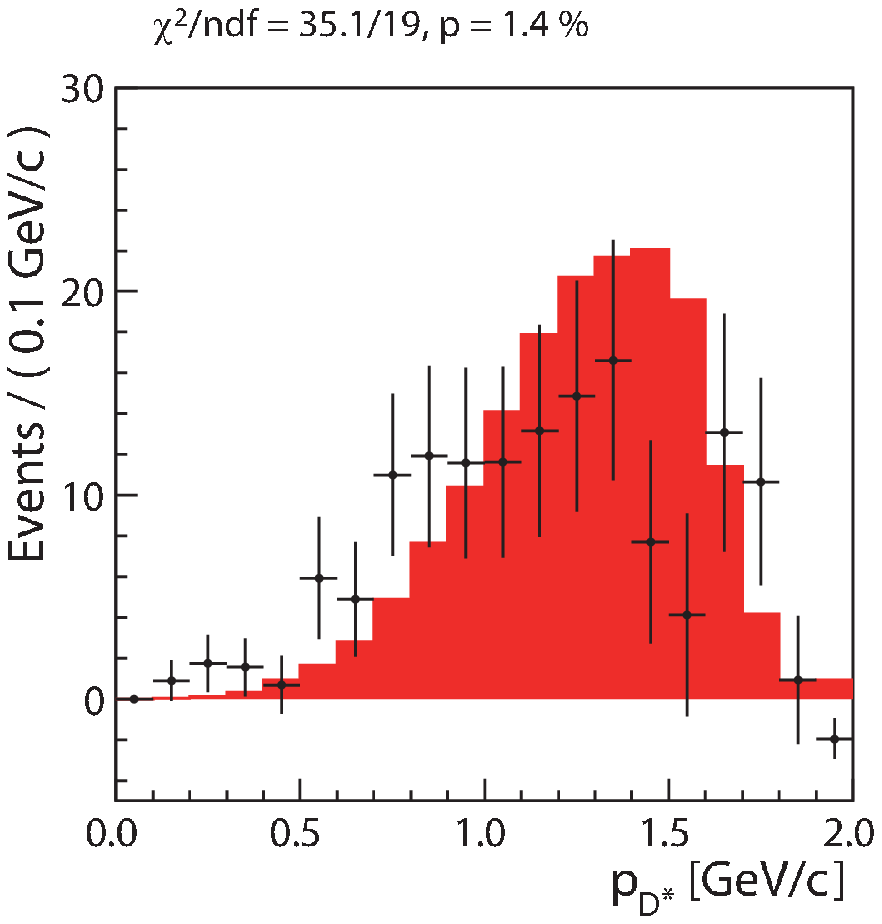}
\label{fig:kinematics_dstr_R2LQ}}
\subfigure{
\includegraphics*[width=5cm]{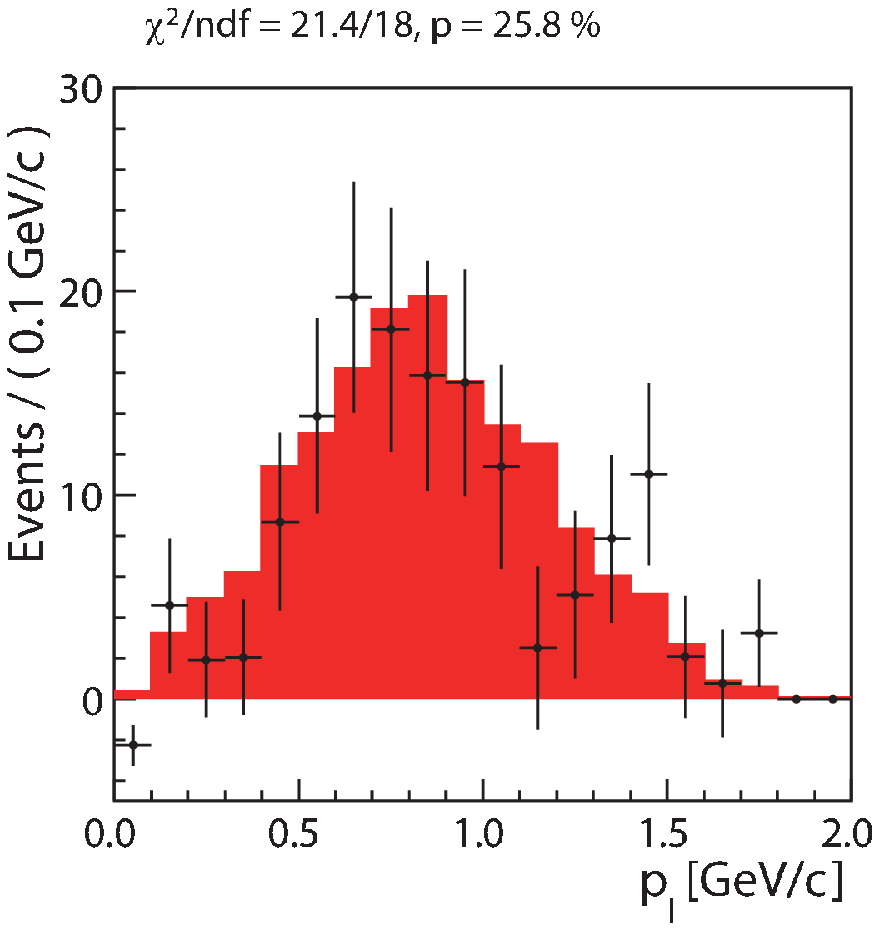}
\label{fig:kinematics_lepton_SM}}
\subfigure{
\includegraphics*[width=5cm]{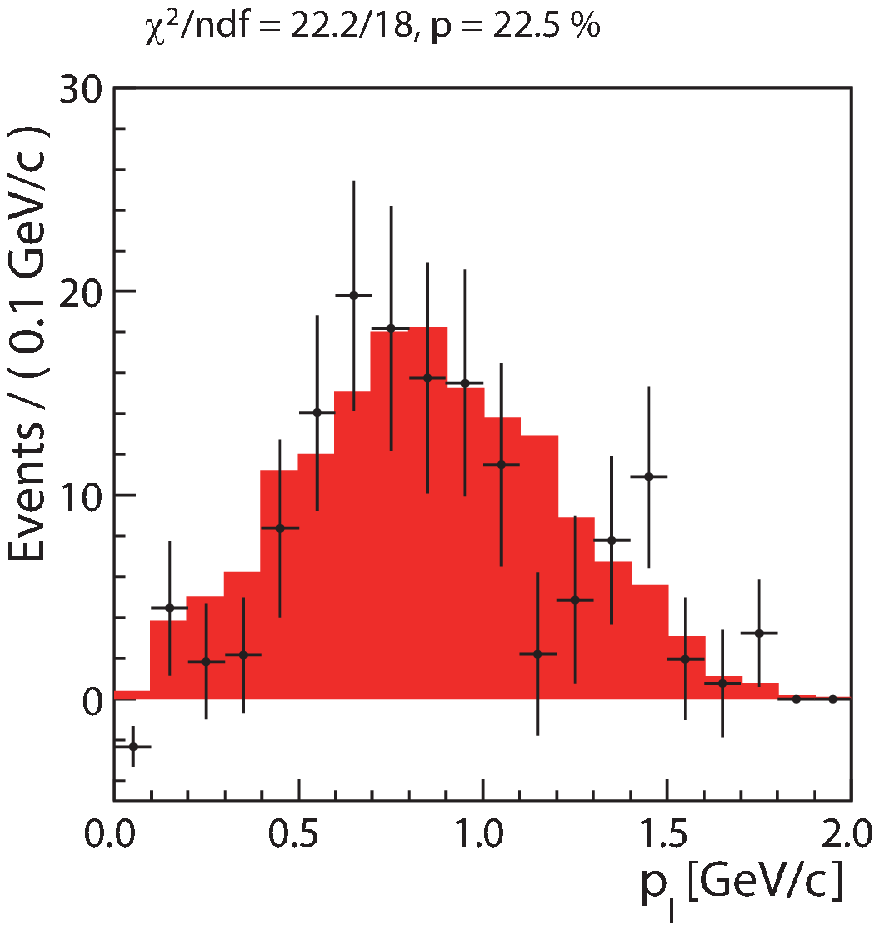}
\label{fig:kinematics_lepton_2hdmII}}
\subfigure{
\includegraphics*[width=5cm]{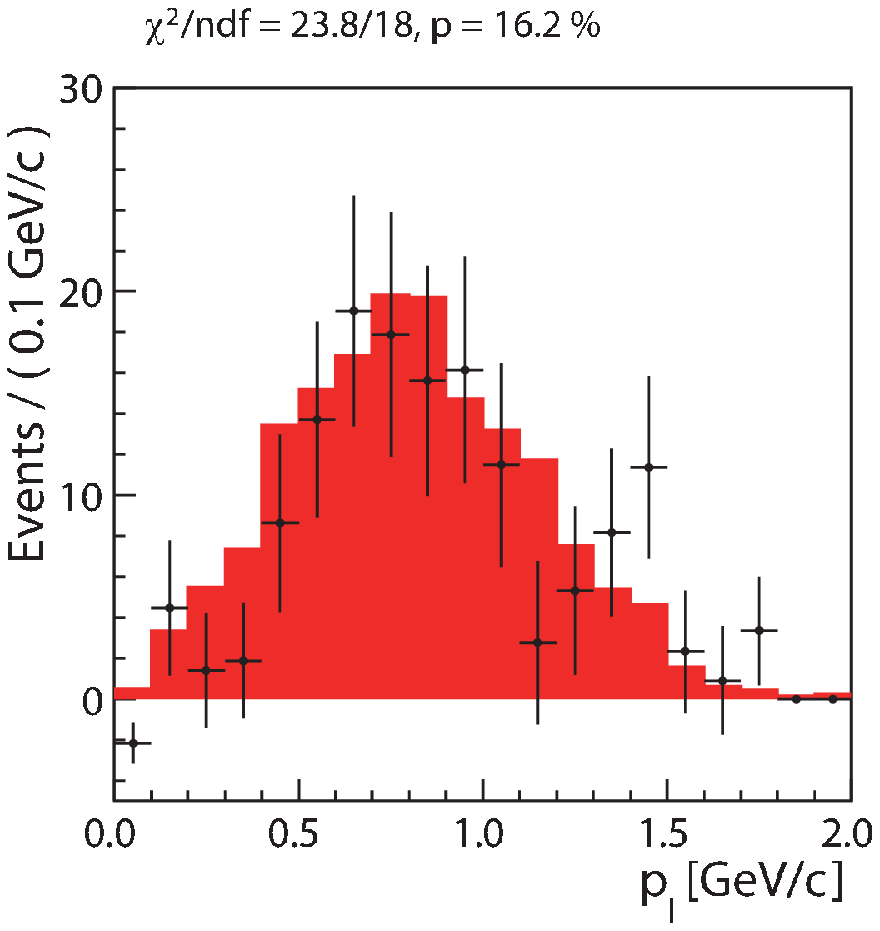}
\label{fig:kinematics_lepton_R2LQ}}
\caption{Background-subtracted momentum distributions of $D^*$ (top) and $\ell$ (bottom)
in the region of ${\cal O}_{\mathit{NB}} > 0.8$ and $E_{\rm ECL} < 0.5$ GeV
for (left) the SM, (center) the type-II 2HDM with $\tan \beta / m_{H^+} = 0.7$ GeV$^{-1}$,
and (right) $R_2$-type leptoquark model with $C_T = +0.36$.
The points and the shaded histograms correspond to the measured and expected distributions, respectively.
The expected distributions are normalized to the number of detected events.
}
\label{fig:kinematics}
\end{figure*}

In Refs.~\cite{BABAR_HAD_NEW} and \cite{BELLE_HAD_NEW},
the $q^2 \equiv (p_B - p_{D^*})^2$ spectra are examined in order to
study the effects of new physics beyond the SM.
Since $q^2$ cannot be calculated here due to the neutrino in the decay of the $B_{\rm tag}$,
we use instead the momenta of the $D^*$ and the $\ell$ in $B_{\rm sig}$ at the $\Upsilon(4S)$ rest frame.
%instead of $q^2$.
Figure~\ref{fig:kinematics} shows the momentum distributions of the background-subtracted data
in the region of ${\cal O}_{\mathit{NB}} > 0.8$ and $E_{\rm ECL} < 0.5$ GeV
for the SM, type-II 2HDM with $\tan \beta / m_{H^+} = 0.7$ GeV$^{-1}$,
and the $R_2$-type leptoquark model with $C_T = +0.36$.
The PDF shapes of background events are taken from MC simulation
and normalized to the yields obtained by the fitting.
Table~\ref{tab:p_values} shows $p$ values for all scenarios,
where we include only the statistical uncertainty.
We find our data are compatible with the SM
and additional contributions from scalar and vector operators;
large additional contributions from tensor operator or the $R_2$- and $S_1$-type leptoquark models are disfavored.

\begin{table}[htbp]
\caption{$p$ values for each scenario
from the momentum distributions of the $D^*$ or the lepton on the signal side in the $\Upsilon(4S) $ rest frame,
where we include only the statistical uncertainty.}
\begin{center}
\begin{tabular}{c|c|c|c} \hline \hline
                       &                                         & \multicolumn{2}{c}{$p$ values [\%]} \\ \cline{3-4}
Model or operator     & Parameter                              & $p_{D^*}$ & $p_{\ell}$ \\ \hline
SM                     &                                         & 37.6    & 25.8 \\
Type-II 2HDM           & $\frac{\tan \beta}{m_{H^+}} = 0.7$ GeV$^{-1}$ & 37.9    & 22.5 \\
%${\cal O}_{V_1}$       & $C_{V_1} = -2.10$                       & 38.8    & 26.2 \\
${\cal O}_{V_2}$       & $C_{V_2} = +1.88$                       & 24.1    & 18.6 \\
${\cal O}_{T}$         & $C_{T}   = +0.36$                       &  0.9    & 19.2 \\
$R_2$-type leptoquark model & $C_T = +0.36$                           &  1.4    & 16.2 \\
$S_1$-type leptoquark model & $C_T = +0.26$                           &  1.1    & 15.4 \\
\hline \hline
\end{tabular}
\label{tab:p_values}
\end{center}
\end{table}

\section{CONCLUSION}
In conclusion,
we report the first measurement of ${\cal R}(D^*)$
with a semileptonic tagging method
using a data sample containing $772 \times 10^6 B\bar{B}$ pairs collected with the Belle detector.
The result is
\begin{eqnarray}
{\cal R}(D^*) &=& 0.302 \pm 0.030 \pm 0.011,
\end{eqnarray}
which is within $1.6 \sigma$ of the SM prediction including systematic uncertainties,
and is in good agreement with other measurements by Belle~\cite{BELLE_INCLUSIVE_OBSERVATION,BELLE_INCLUSIVE, BELLE_HAD_NEW},
\mbox{\sl B\hspace{-0.4em} {\small\sl A}\hspace{-0.37em} \sl B\hspace{-0.4em}
{\small\sl A\hspace{-0.02em}R}}~\cite{BABAR_HAD_NEW},
and LHCb~\cite{LHCB_RESULT}.
The result is statistically independent of earlier Belle measurements.
We investigate the compatibility
of the data samples with new physics in a model-independent method by adding the operators one by one.
We also study two types of leptoquark models.
We find our data allow the additional contributions from scalar and vector operators
while disfavoring large additional contributions from
a tensor operator with $+0.34 < C_T < +0.39$,
an $R_2$-type leptoquark model with $+0.34 < C_T < +0.38$,
or
an $S_1$-type leptoquark model with $+0.22 < C_T < +0.28$,
when considering the impact on the decay kinematics.

\section{ACKNOWLEDGEMENTS}
%----------- Long version, for most papers ----------- 
We thank Y.~Sakaki, R.~Watanabe, and M.~Tanaka for their invaluable suggestions.
This work was supported in part by
a Grant-in-Aid for JSPS Fellows (No.13J03438)
and 
a Grant-in-Aid for Scientific Research (S) ``Probing New Physics with Tau-Lepton'' (No.26220706).
We thank the KEKB group for the excellent operation of the
accelerator; the KEK cryogenics group for the efficient
operation of the solenoid; and the KEK computer group,
the National Institute of Informatics, and the 
PNNL/EMSL computing group for valuable computing
and SINET4 network support.  We acknowledge support from
the Ministry of Education, Culture, Sports, Science, and
Technology (MEXT) of Japan, the Japan Society for the 
Promotion of Science (JSPS), and the Tau-Lepton Physics 
Research Center of Nagoya University; 
the Australian Research Council;
Austrian Science Fund under Grant No.~P 22742-N16 and P 26794-N20;
the National Natural Science Foundation of China under Contracts 
No.~10575109, No.~10775142, No.~10875115, No.~11175187, No.~11475187
and No.~11575017;
the Chinese Academy of Science Center for Excellence in Particle Physics; 
the Ministry of Education, Youth and Sports of the Czech
Republic under Contract No.~LG14034;
the Carl Zeiss Foundation, the Deutsche Forschungsgemeinschaft, the
Excellence Cluster Universe, and the VolkswagenStiftung;
the Department of Science and Technology of India; 
the Istituto Nazionale di Fisica Nucleare of Italy; 
the WCU program of the Ministry of Education, National Research Foundation (NRF) 
of Korea Grants No.~2011-0029457,  No.~2012-0008143,  
No.~2012R1A1A2008330, No.~2013R1A1A3007772, No.~2014R1A2A2A01005286, 
No.~2014R1A2A2A01002734, No.~2015R1A2A2A01003280, No. 2015H1A2A1033649;
the Basic Research Lab program under NRF Grant No.~KRF-2011-0020333,
Center for Korean J-PARC Users, No.~NRF-2013K1A3A7A06056592; 
the Brain Korea 21-Plus program and Radiation Science Research Institute;
the Polish Ministry of Science and Higher Education and 
the National Science Center;
the Ministry of Education and Science of the Russian Federation and
the Russian Foundation for Basic Research;
the Slovenian Research Agency;
Ikerbasque, Basque Foundation for Science and
the Euskal Herriko Unibertsitatea (UPV/EHU) under program UFI 11/55 (Spain);
the Swiss National Science Foundation; 
the Ministry of Education and the Ministry of Science and Technology of Taiwan;
and the U.S.\ Department of Energy and the National Science Foundation.
This work is supported by a Grant-in-Aid from MEXT for 
Science Research in a Priority Area (``New Development of 
Flavor Physics'') and from JSPS for Creative Scientific 
Research (``Evolution of Tau-lepton Physics'').

\clearpage

%SUPPLEMENTAL MATERIAL
\section{SUPPLEMENTAL MATERIAL}

\begin{figure*}[htb]
\centering
%\vspace{-25mm}
  \includegraphics[width=11.0cm]{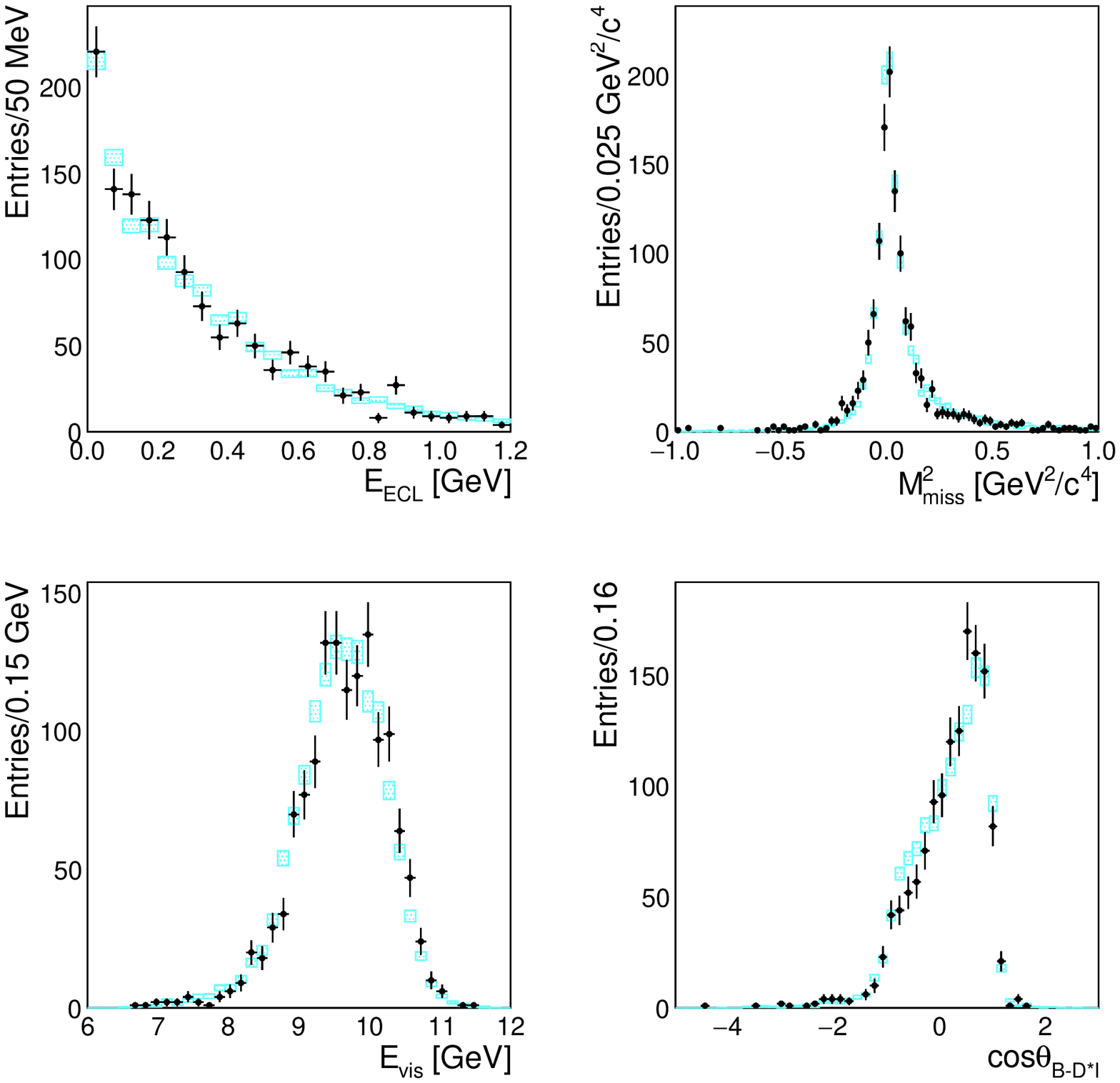}
  \caption{Distributions of $E_{\rm ECL}$ (top left), $M_{\rm miss}^2$ (top right), $E_{\rm vis}$ (bottom left),
  and $\cos \theta_{B \mathchar`-  D^* \ell}$ (bottom right) in the hadronic tagged sample.
  The signal side is reconstructed from the decay $\bar{B}^0 \rightarrow D^{*+} \ell^- \bar{\nu}_{\ell}$
  followed by $D^{*+} \rightarrow D^0 \pi^+$.
  The black dots with error bars show the data
  and the light blue rectangles show the normalized MC samples.}
  \label{fig:hadtag}
\end{figure*}

\begin{figure*}[htb]
\centering
%\vspace{-25mm}
\subfigure[${\cal O}_{\mathit{NB}}$ distribution.]{
\includegraphics*[width=5.3cm]{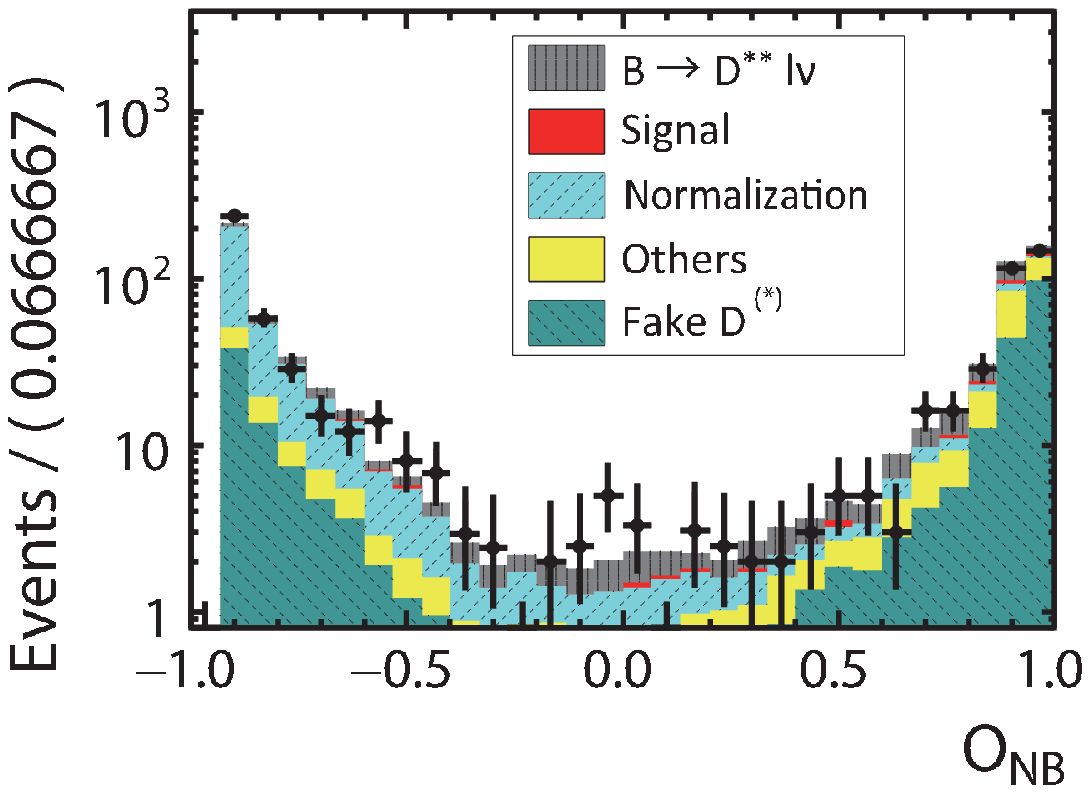}
\label{fig:ctrl_fit_nn}}
\subfigure[$E'_{\rm ECL}$ distribution with $B \rightarrow D^{**} \ell \nu_{\ell}$-enhanced ${\cal O}_{\mathit{NB}}$ region (${\cal O}_{\mathit{NB}} > 0.0$).]{
\includegraphics*[width=5.3cm]{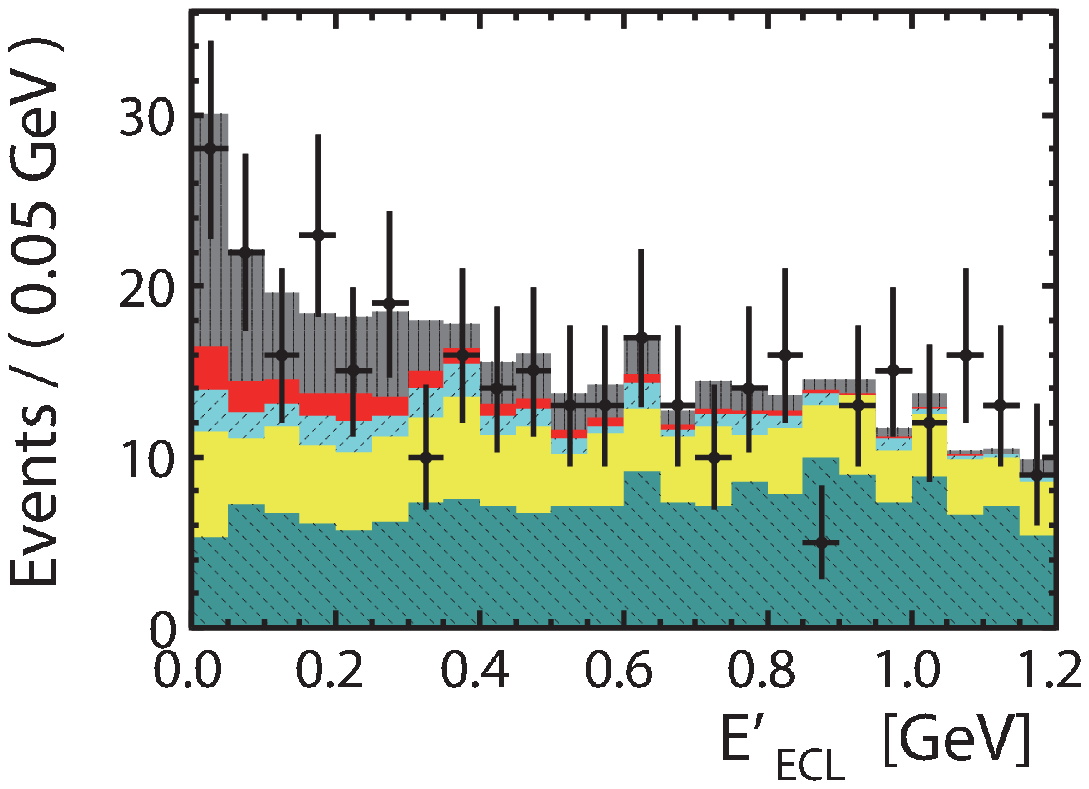}
\label{fig:ctrl_fit_eecl_sig}}
\subfigure[$E'_{\rm ECL}$ distribution with normalization-enhanced ${\cal O}_{\mathit{NB}}$ region (${\cal O}_{\mathit{NB}} < 0.0$).]{
\includegraphics*[width=5.3cm]{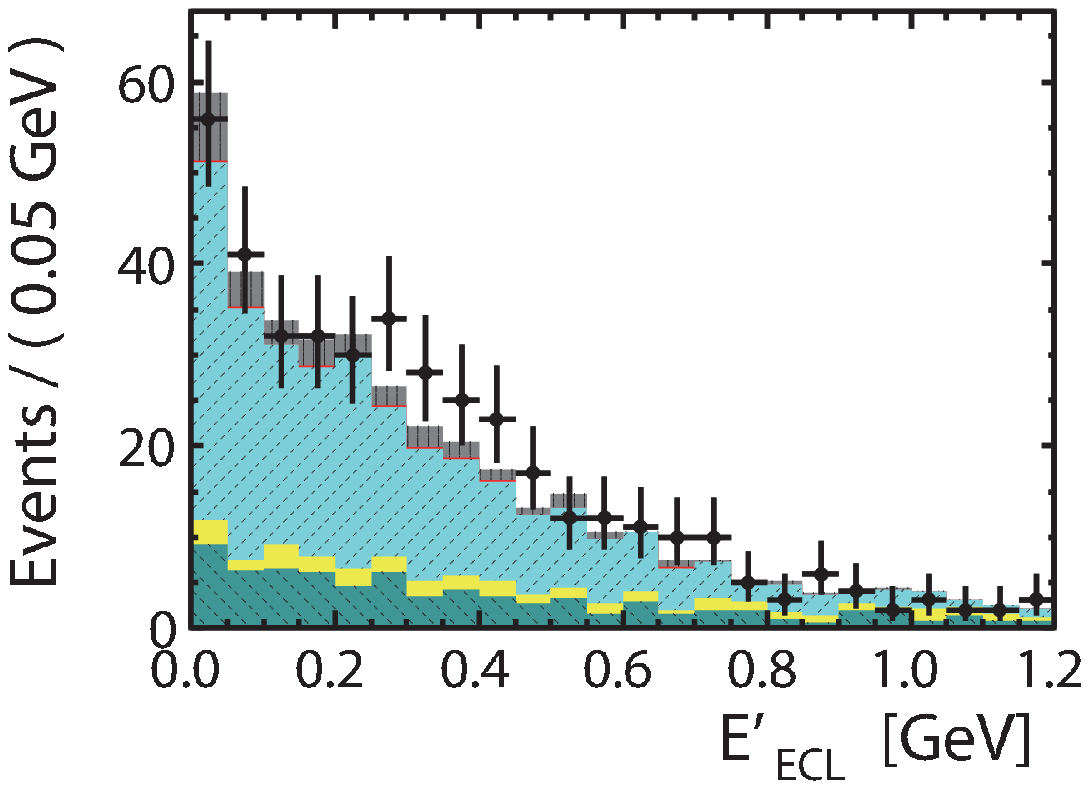}
\label{fig:ctrl_fit_eecl_norm}}
\caption{
Projection of the fit results of control samples with data points overlaid.
In the control samples,
we require $\pi^0$ in addition to the nominal event selection.
$E'_{\rm ECL}$ is defined as $E_{\rm ECL}$ excluded from energy deposit from additional $\pi^0$.
}
\label{fig:ctrl_fit_supp}
\end{figure*}

\begin{figure*}[htb]
\centering
%\vspace{-25mm}
\subfigure[Type-II 2HDM.]{
\includegraphics*[width=6cm]{npcurve_2hdmII_eff.eps}
\label{fig:eff_ratio_2hdm}}
\subfigure[SM with adding contribution from ${\cal O}_{S_1}$.]{
\includegraphics*[width=6cm]{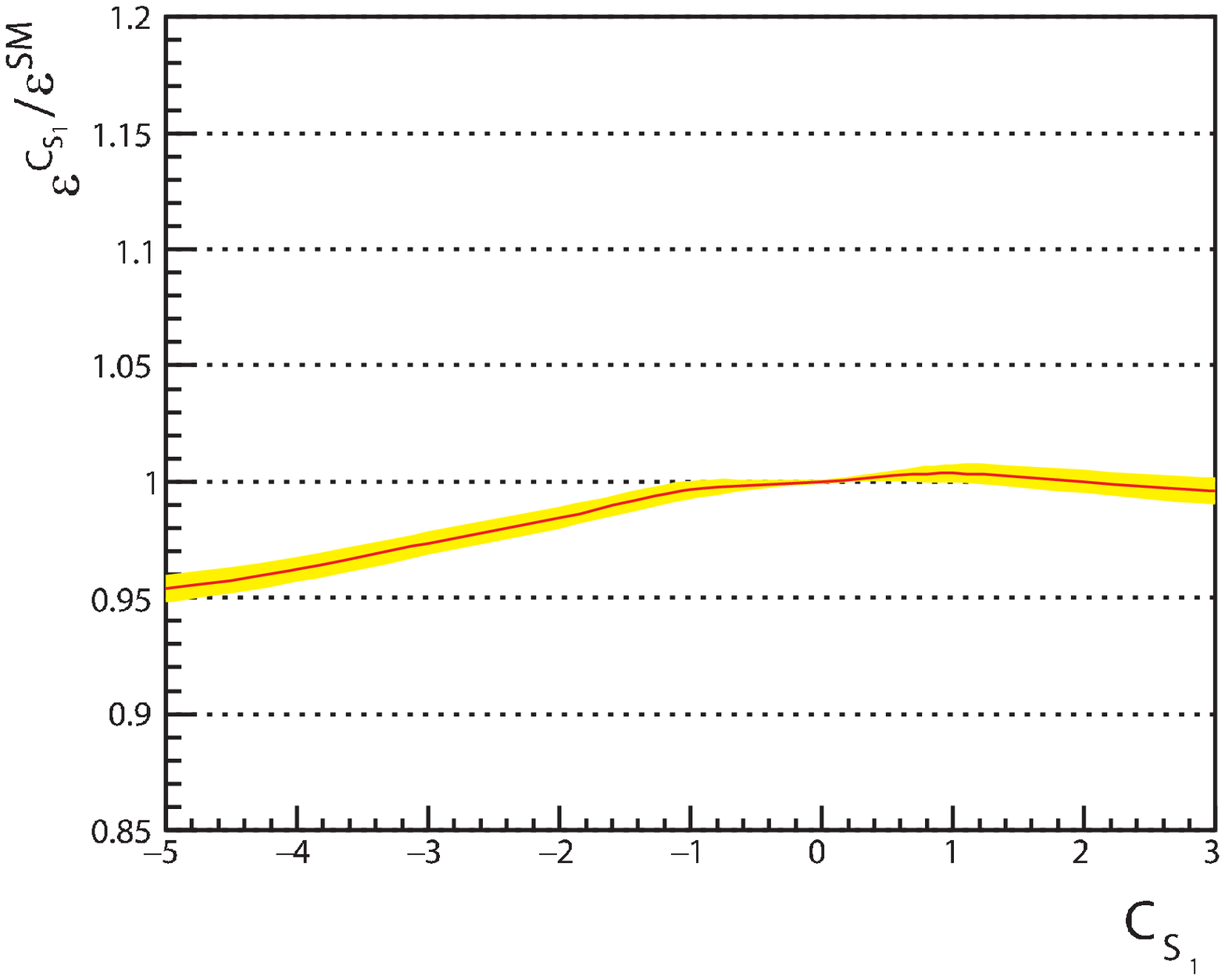}
\label{fig:eff_ratio_OPS1}}
\subfigure[SM with adding contribution from ${\cal O}_{S_2}$.]{
\includegraphics*[width=6cm]{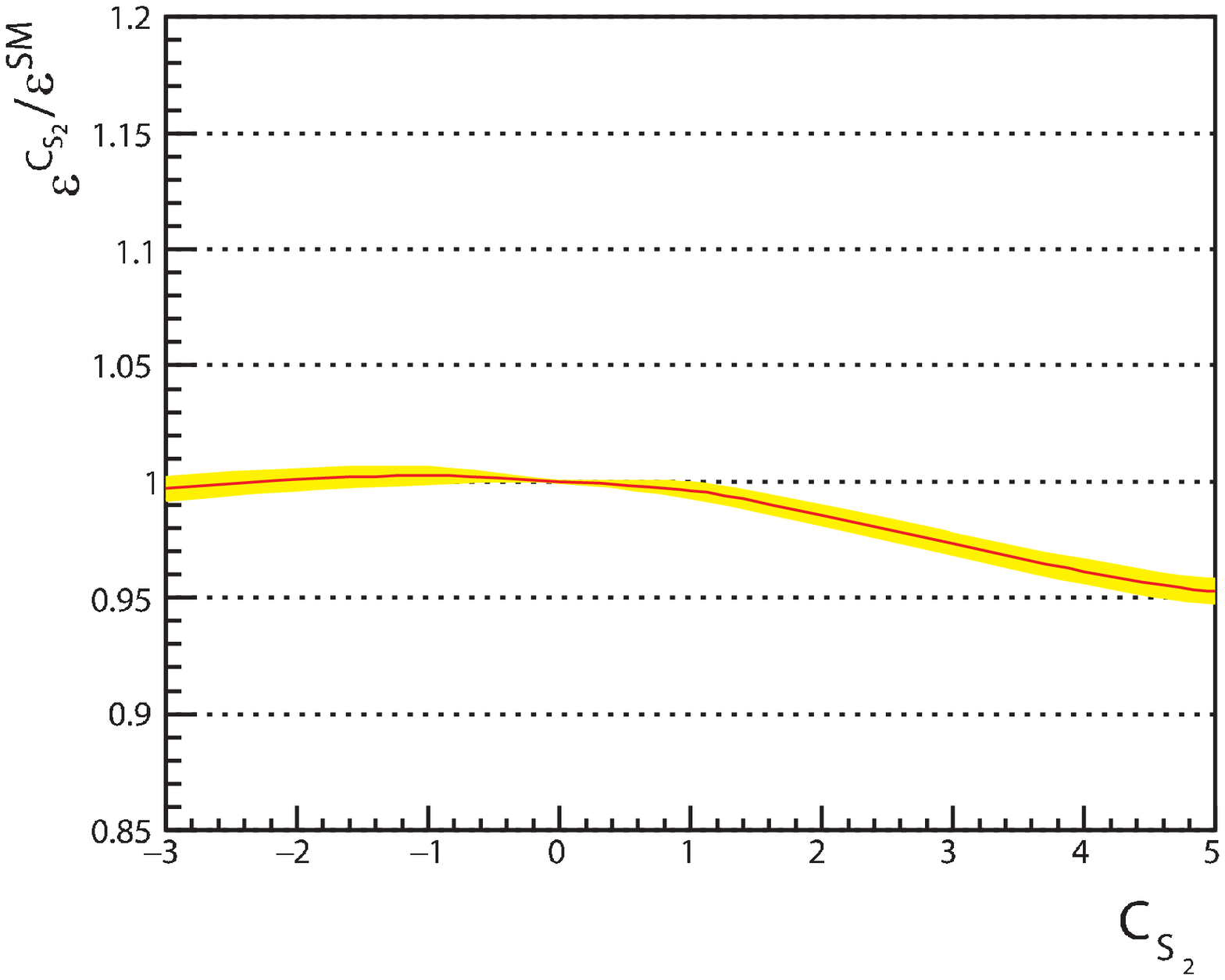}
\label{fig:eff_ratio_OPS2}}
\subfigure[SM with adding contribution from ${\cal O}_{V_1}$.]{
\includegraphics*[width=6cm]{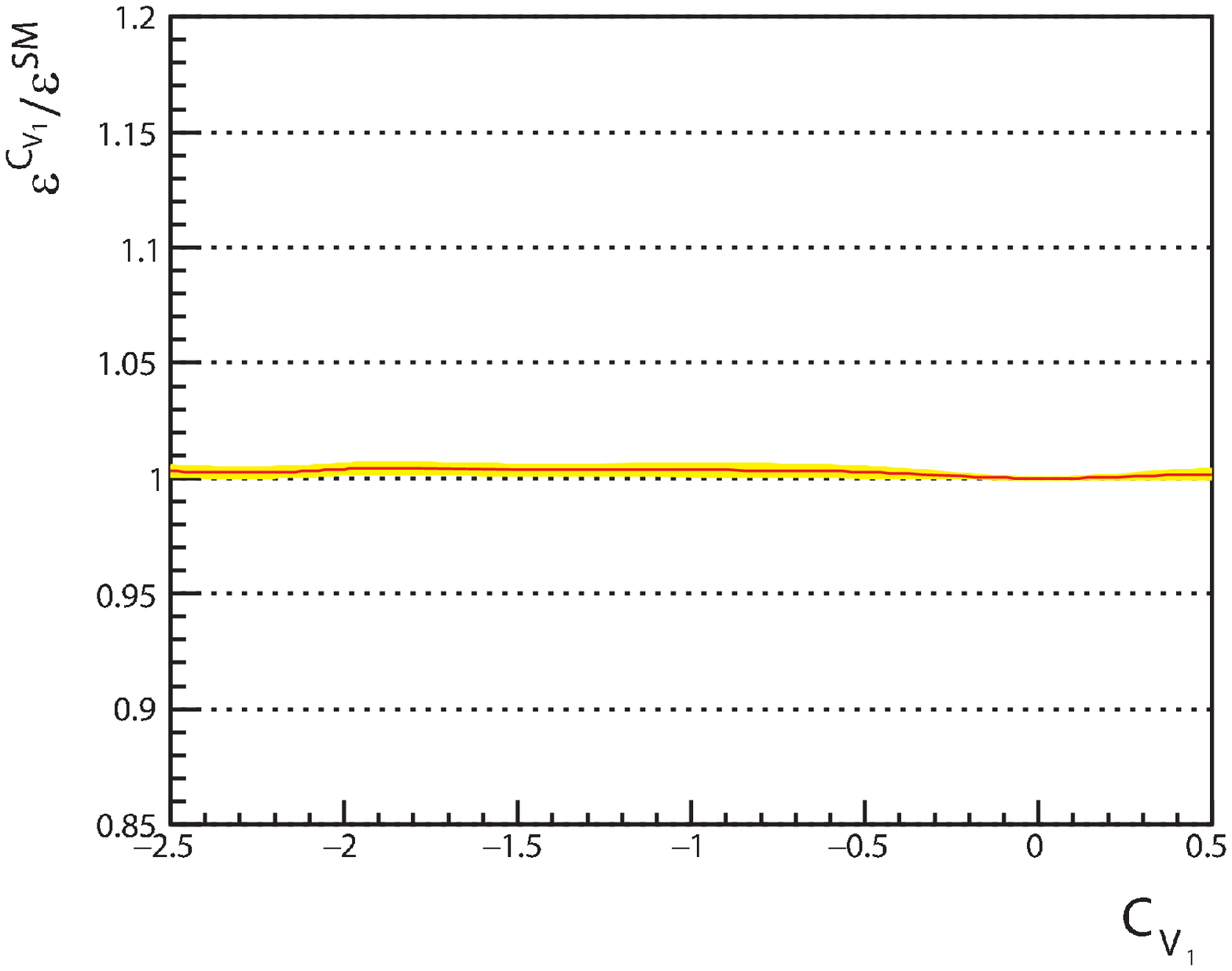}
\label{fig:eff_ratio_OPV1}}
\subfigure[SM with adding contribution from ${\cal O}_{V_2}$.]{
\includegraphics*[width=6cm]{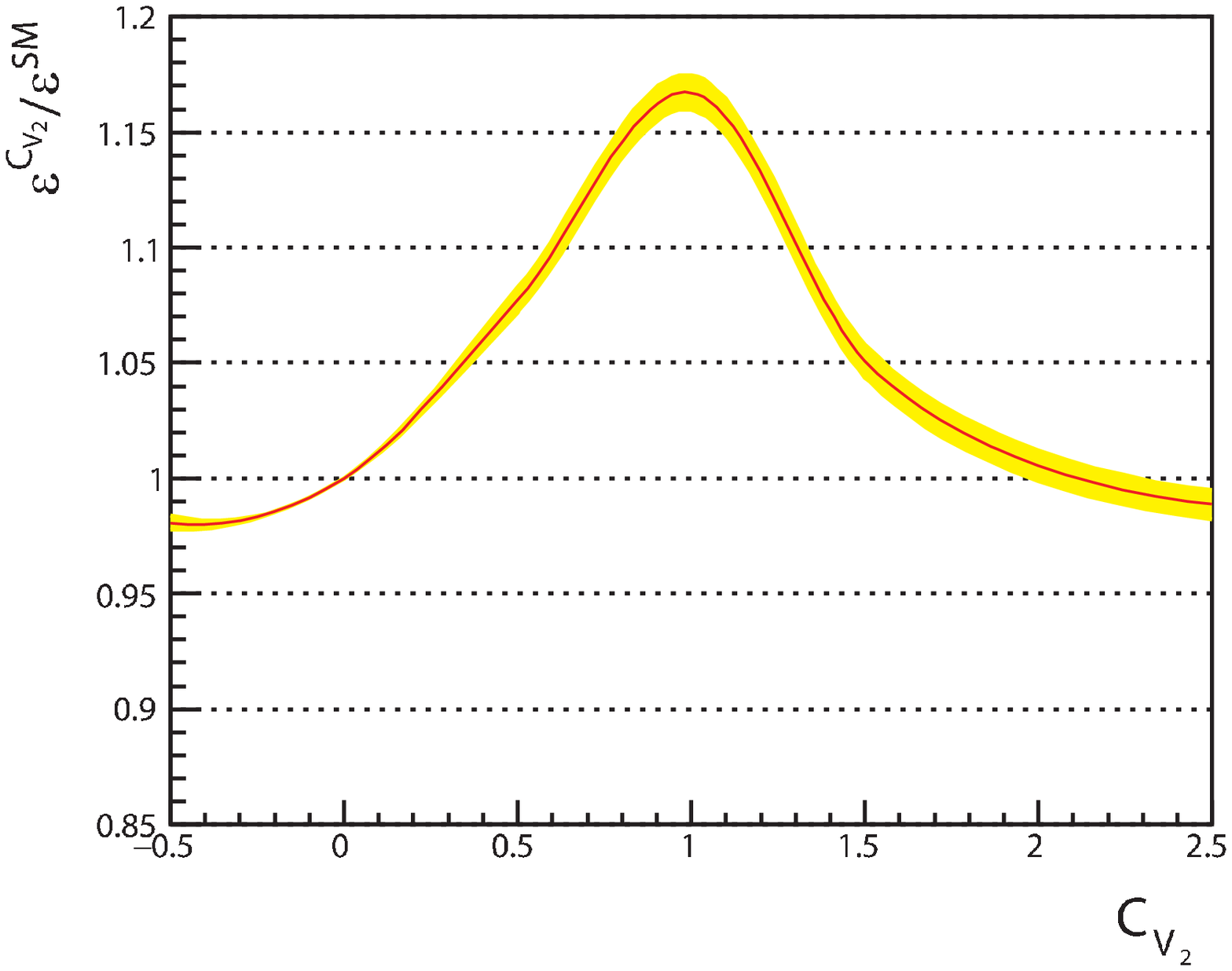}
\label{fig:eff_ratio_OPV2}}
\subfigure[SM with adding contribution from ${\cal O}_{T}$.]{
\includegraphics*[width=6cm]{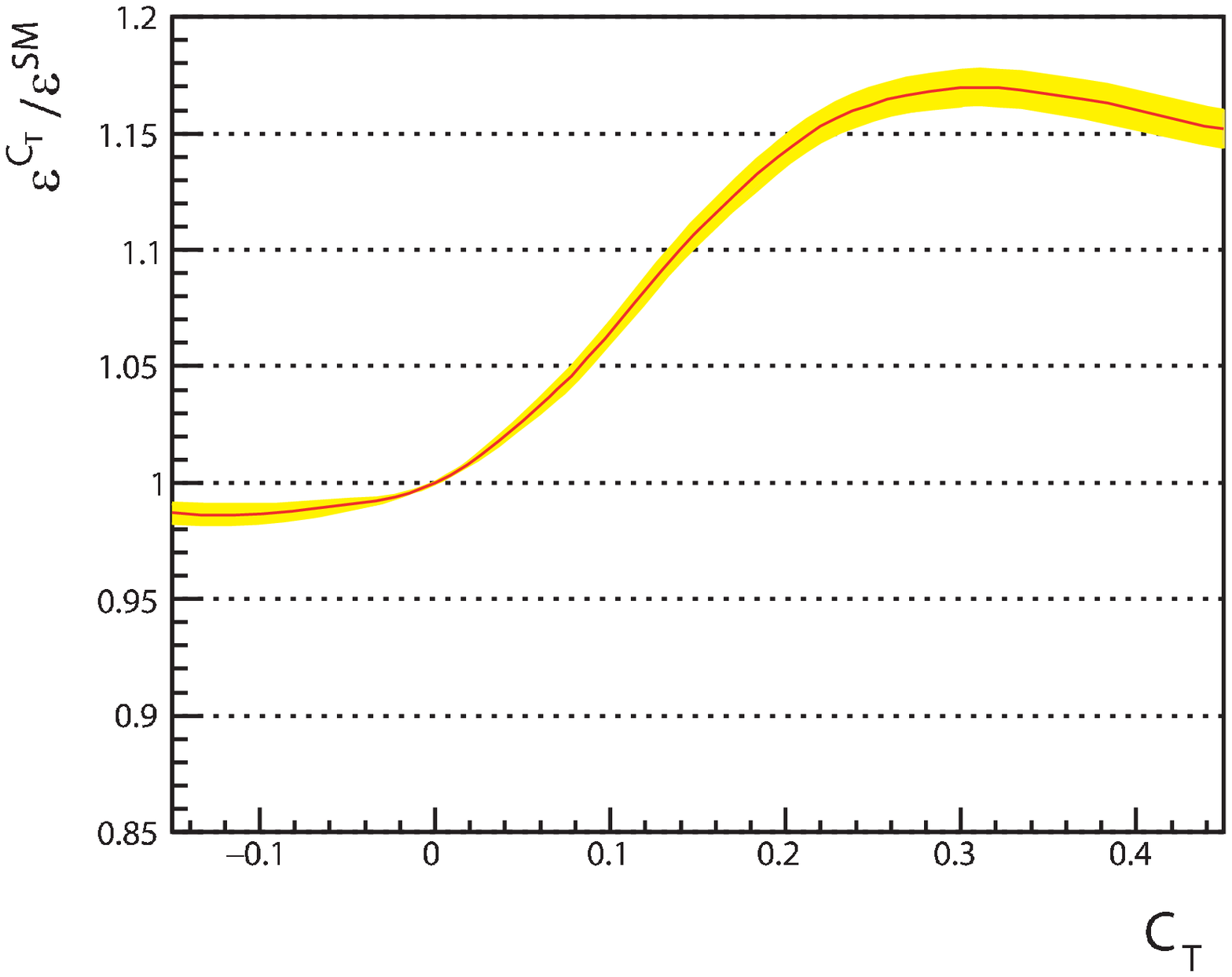}
\label{fig:eff_ratio_OPT}}
\subfigure[$R_2$-type leptoquark model.]{
\includegraphics*[width=6cm]{npcurve_R2LQ_eff.eps}
\label{fig:eff_ratio_R2LQ}}
\subfigure[$S_1$-type leptoquark model.]{
\includegraphics*[width=6cm]{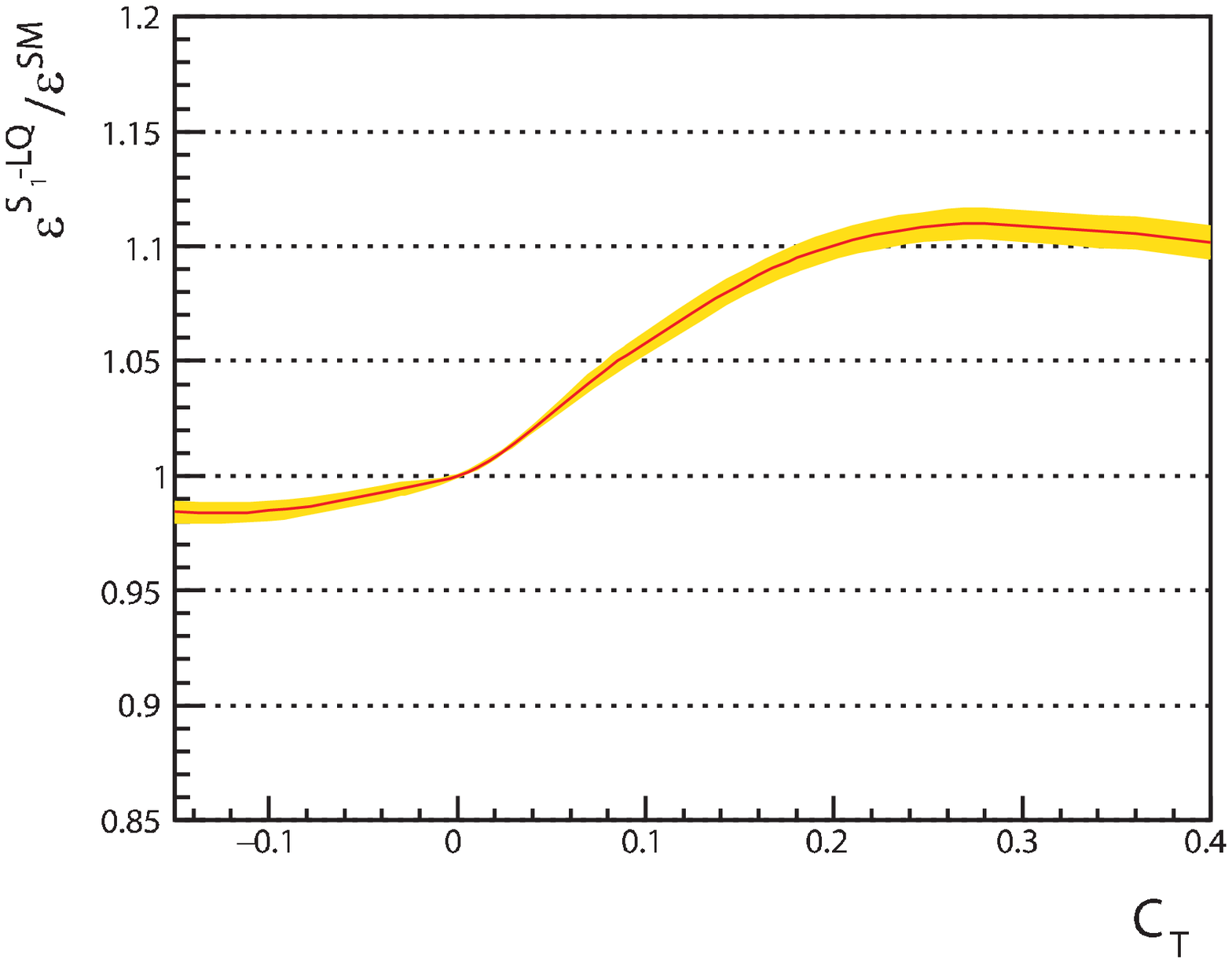}
\label{fig:eff_ratio_S1LQ}}
\caption{
Efficiency with respect to the SM value.
}
\label{fig:npcurve_eff_supp}
\end{figure*}

\begin{figure*}[htb]
\centering
%\vspace{-25mm}
\subfigure[Type-II 2HDM.]{
\includegraphics*[width=6cm]{npcurve_2hdmII_rdstr.eps}
\label{fig:rdstr_2hdm}}
\subfigure[SM with adding contribution from ${\cal O}_{S_1}$.]{
\includegraphics*[width=6cm]{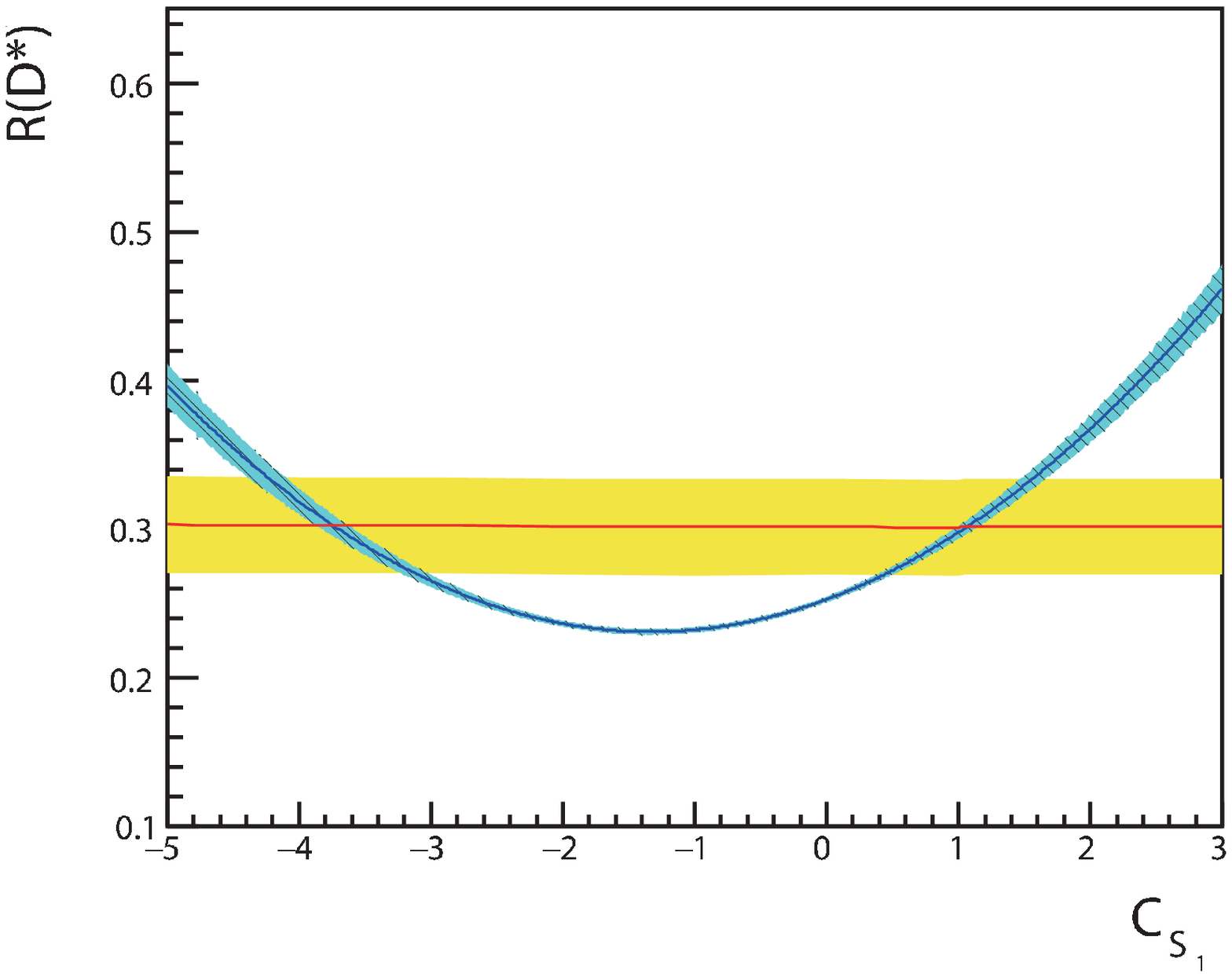}
\label{fig:rdstr_OPS1}}
\subfigure[SM with adding contribution from ${\cal O}_{S_2}$.]{
\includegraphics*[width=6cm]{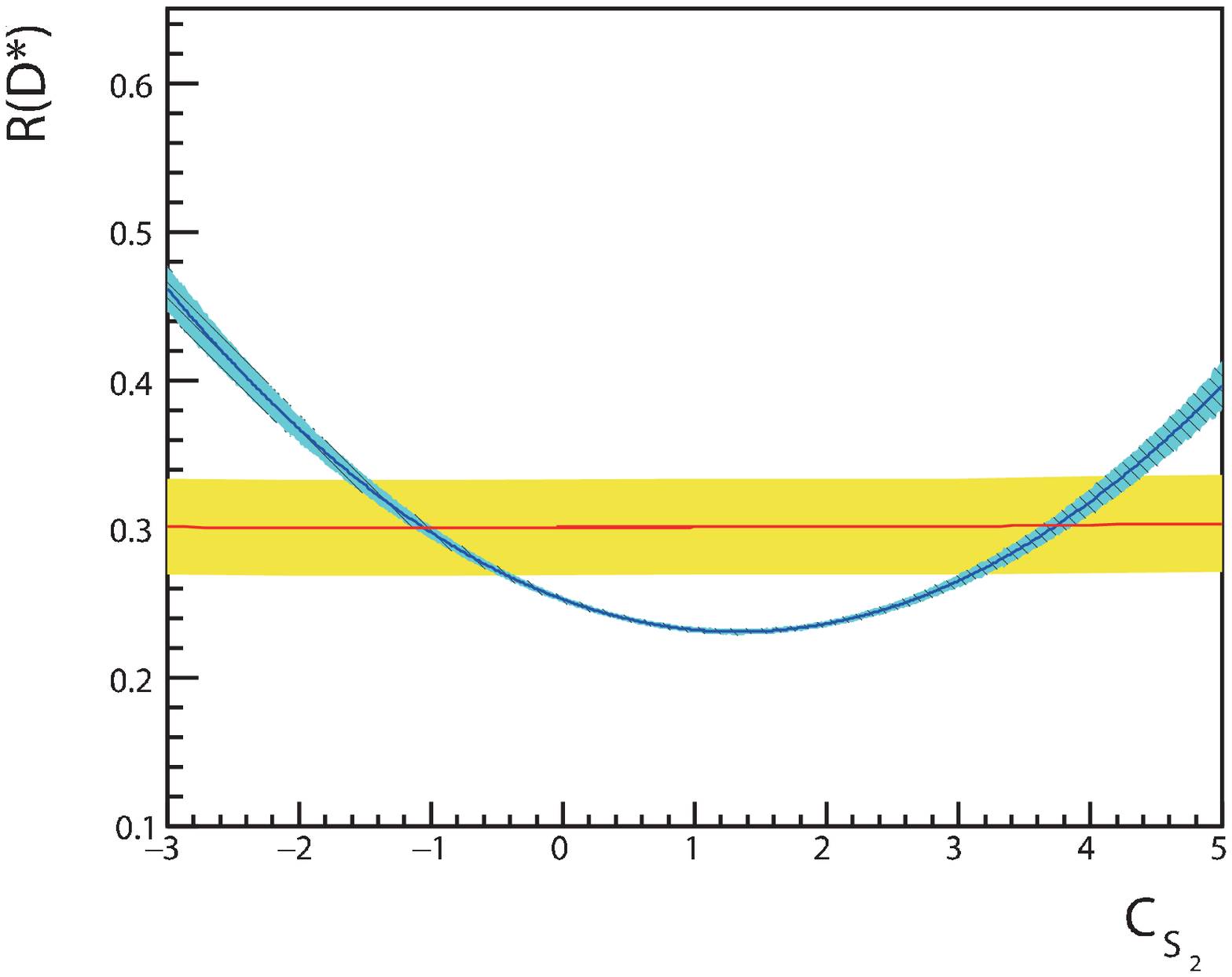}
\label{fig:rdstr_OPS2}}
\subfigure[SM with adding contribution from ${\cal O}_{V_1}$.]{
\includegraphics*[width=6cm]{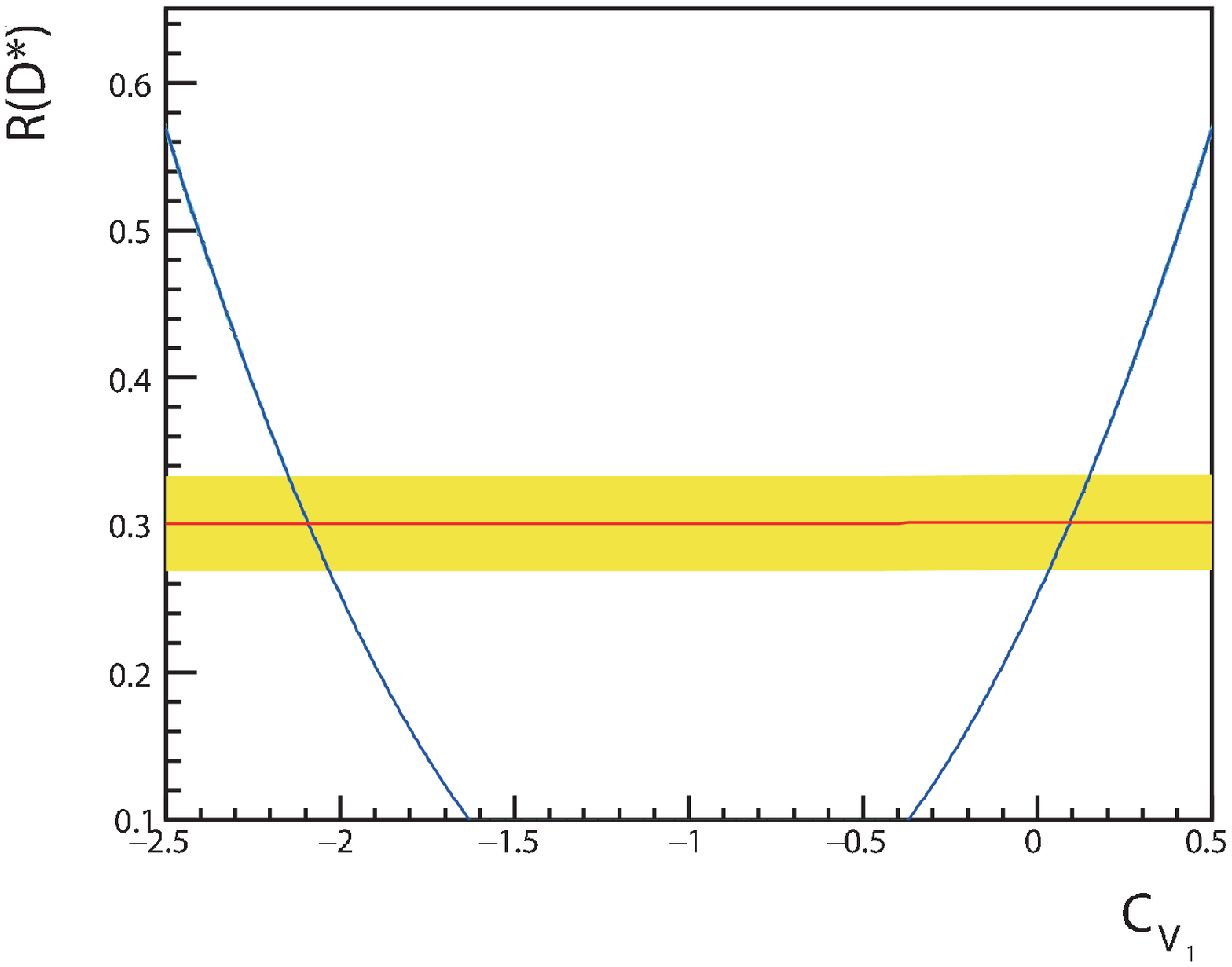}
\label{fig:rdstr_OPV1}}
\subfigure[SM with adding contribution from ${\cal O}_{V_2}$.]{
\includegraphics*[width=6cm]{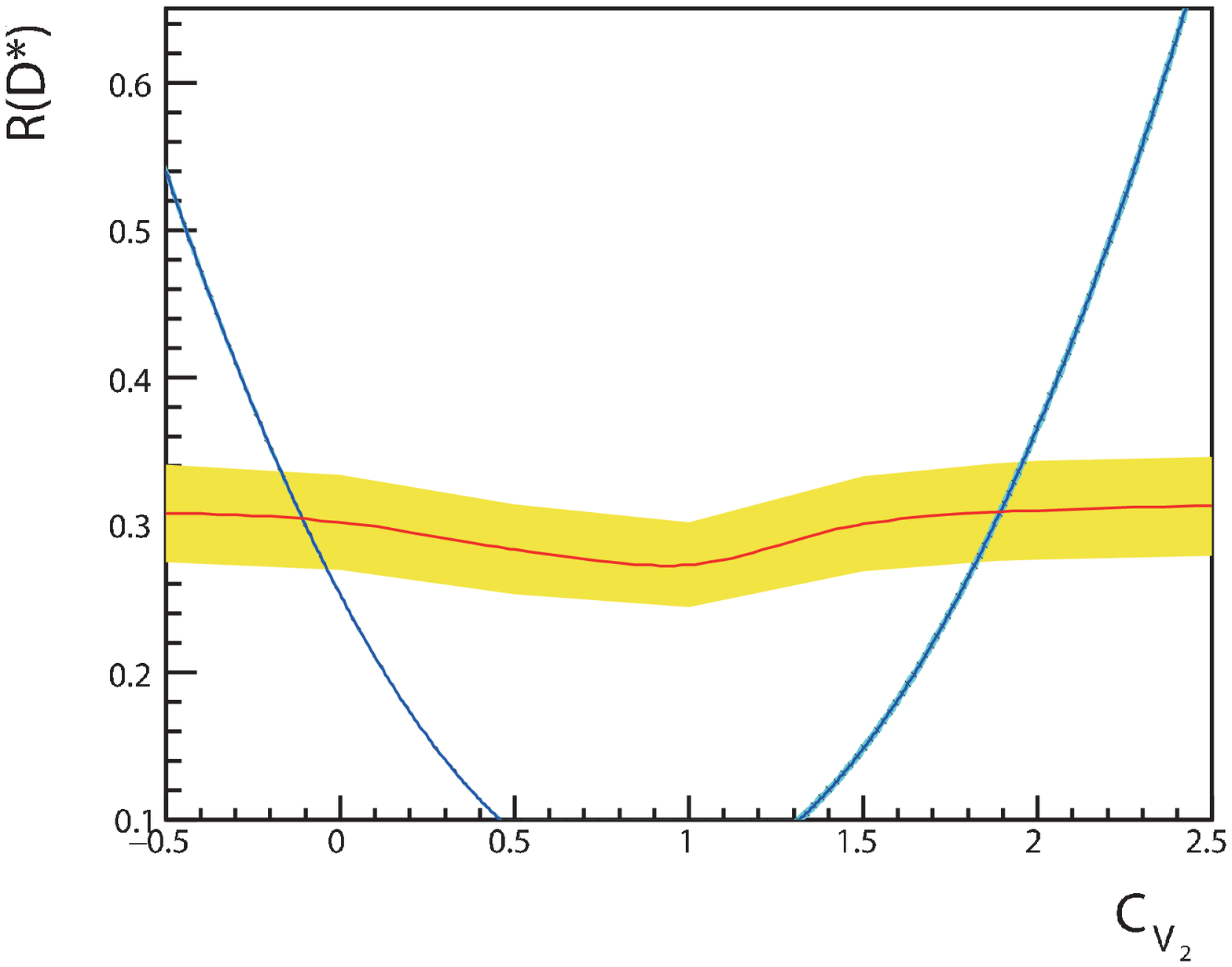}
\label{fig:rdstr_OPV2}}
\subfigure[SM with adding contribution from ${\cal O}_{T}$.]{
\includegraphics*[width=6cm]{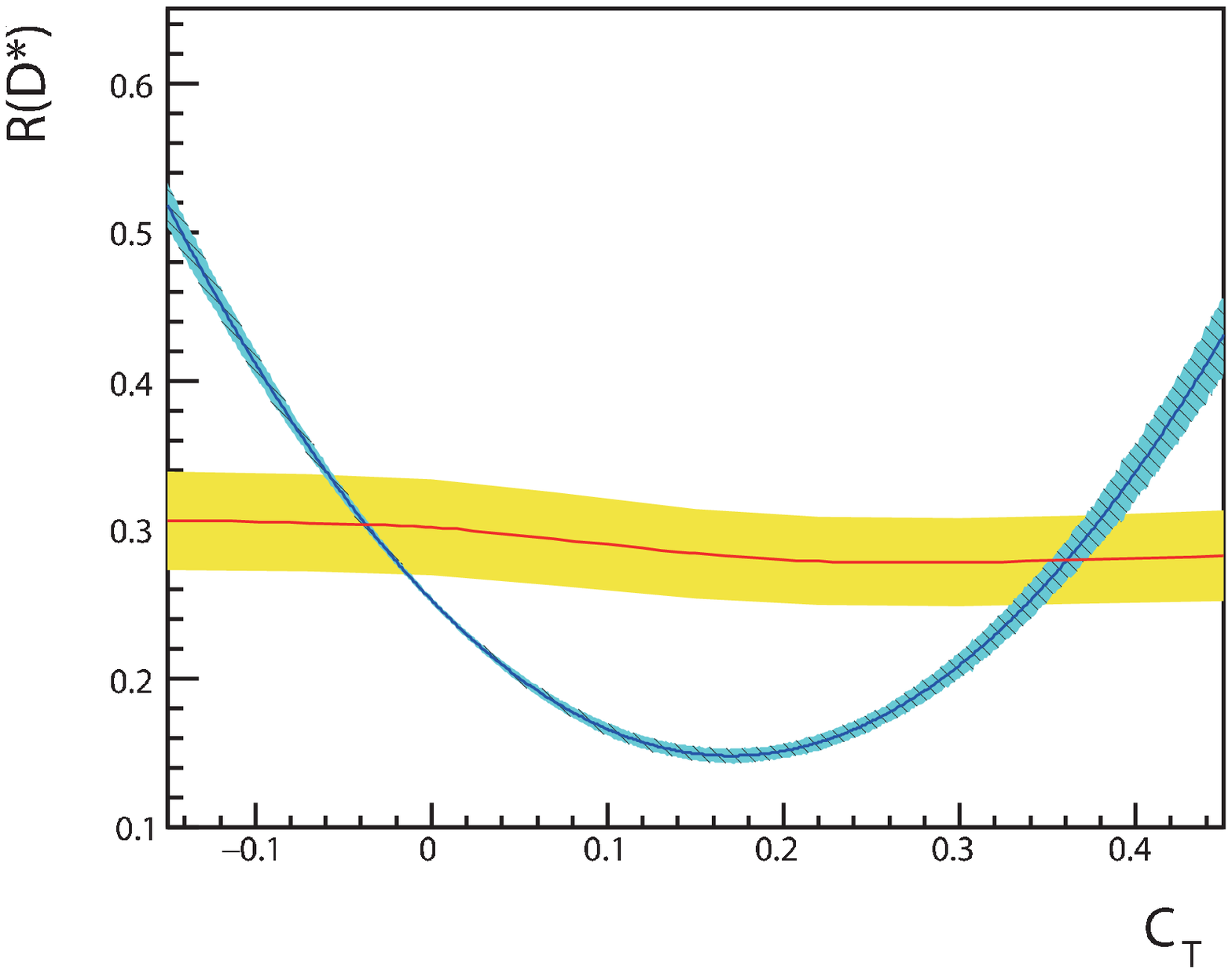}
\label{fig:rdstr_OPT}}
\subfigure[$R_2$-type leptoquark model.]{
\includegraphics*[width=6cm]{npcurve_R2LQ_rdstr.eps}
\label{fig:rdstr_R2LQ}}
\subfigure[$S_1$-type leptoquark model.]{
\includegraphics*[width=6cm]{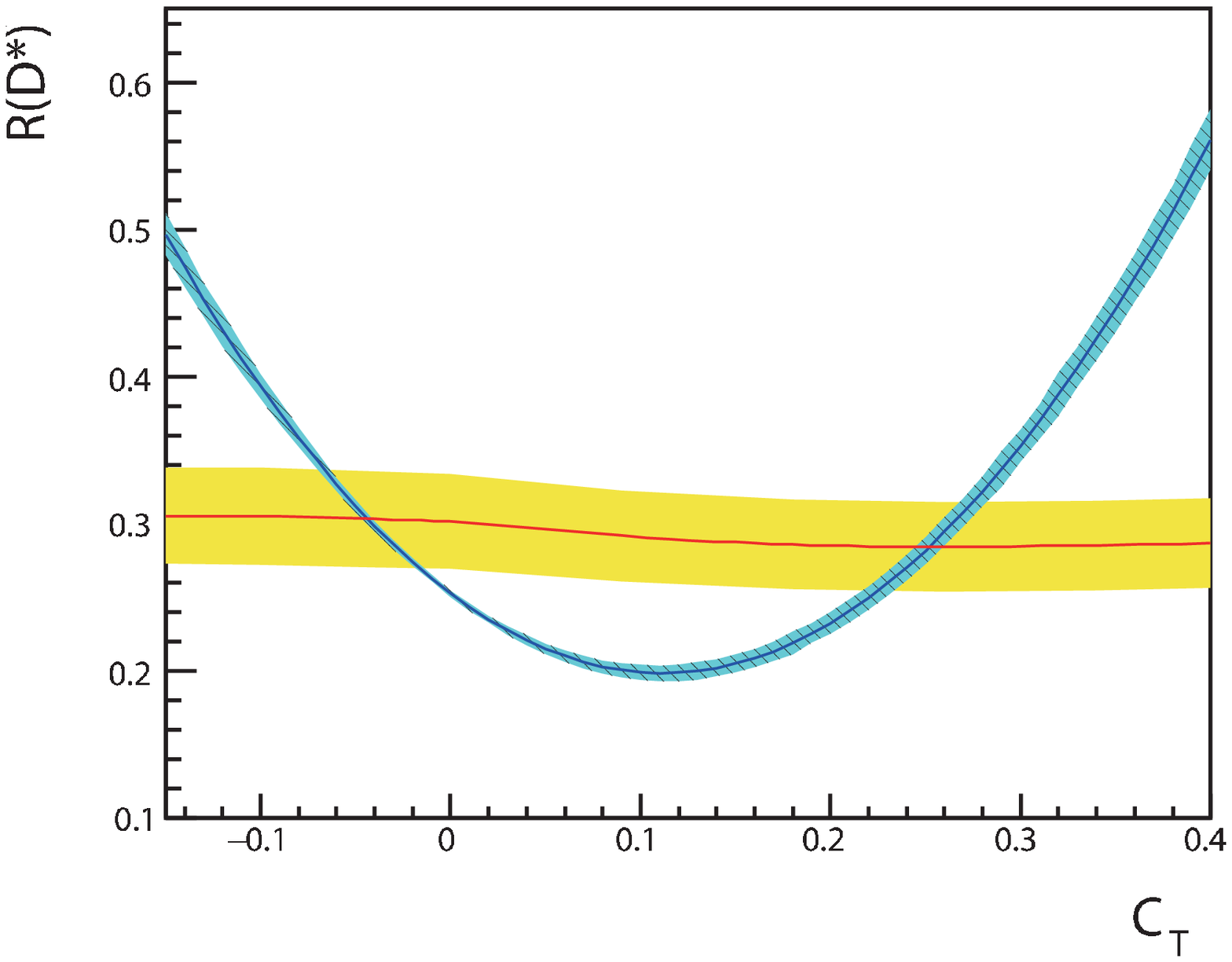}
\label{fig:rdstr_S1LQ}}
\caption{
Measured values of ${\cal R}(D^*)$ and their ($1\sigma$) uncertainties are shown
by solid (red) curve and shaded (yellow) region.
Theoretical predictions and their ($1\sigma$) uncertainties are shown by solid (blue) curve and hatched (light blue) region.
}
\label{fig:rdstr_supp}
\end{figure*}

\begin{figure*}[htb]
\centering
%\vspace{-25mm}
\subfigure[SM.]{
\includegraphics*[width=6.0cm]{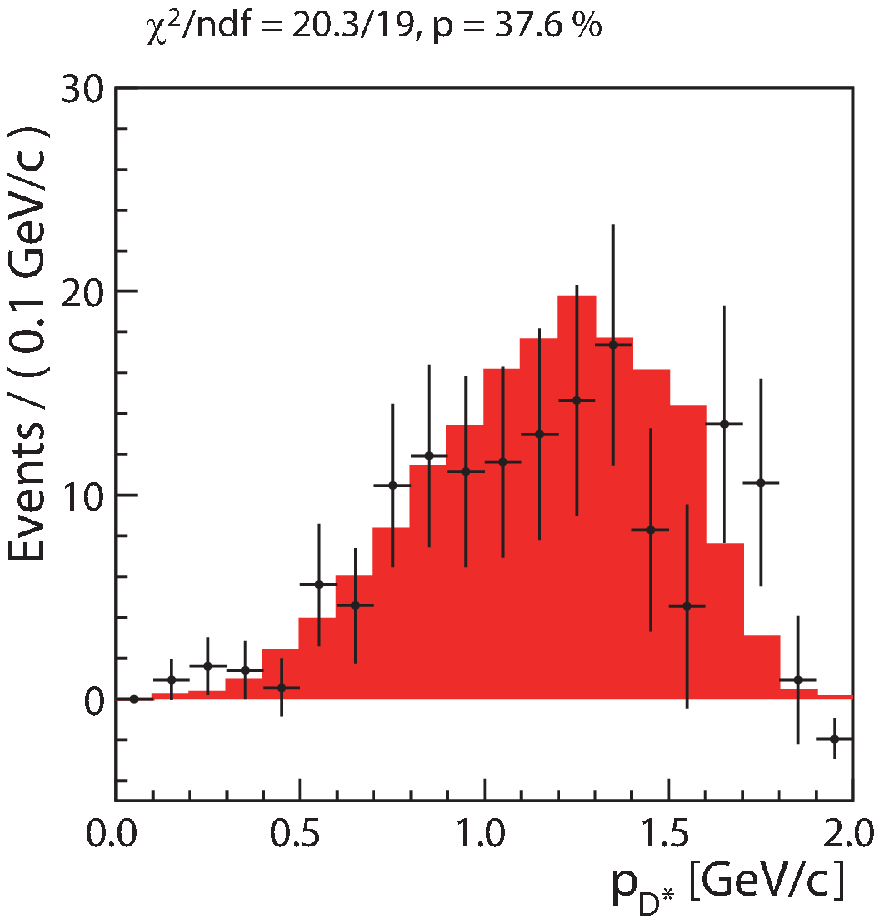}
\label{fig:kinematic_dstr_sm}}
\subfigure[Type-II 2HDM with $\tan \beta / m_{H^+} = 0.7$ GeV$^{-1}$.]{
\includegraphics*[width=6.0cm]{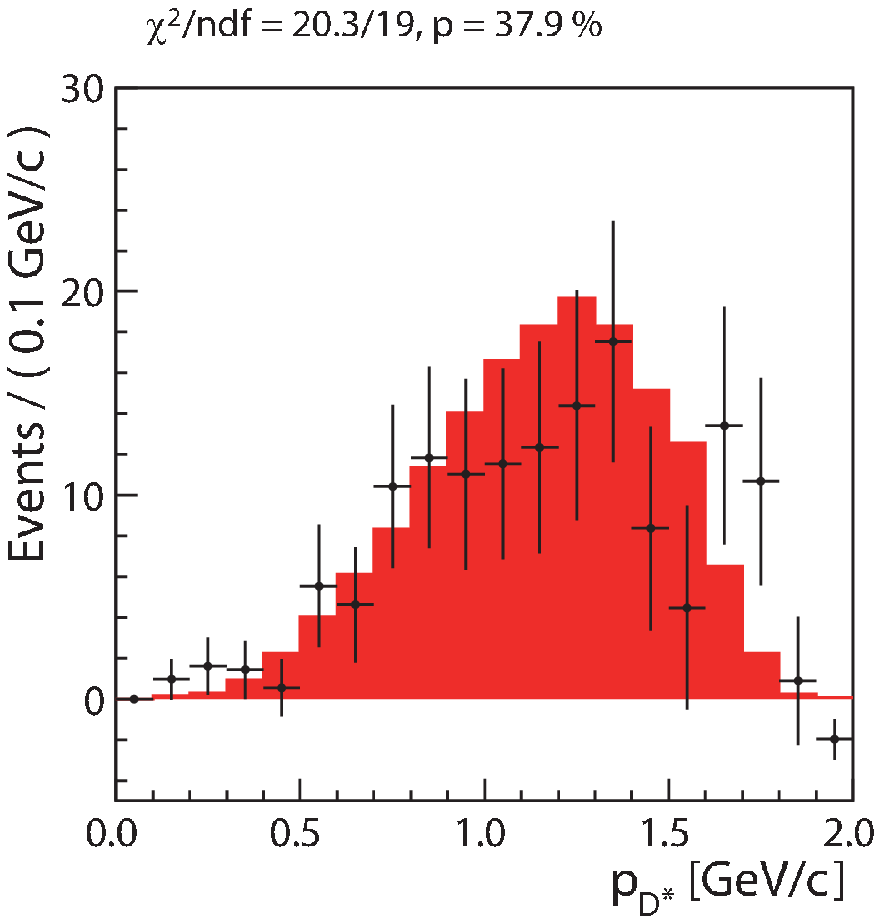}
\label{fig:kinematic_dstr_2hdm}}
%\subfigure[SM with adding contribution from ${\cal O}_{V_1}$ ($C_{V_1} = -2.10$).]{
%\includegraphics*[width=6.0cm]{kinematic_dstr_after_OPV1_m2p10.eps}
%\label{fig:kinematic_dstr_OPV1}}
\subfigure[SM with adding contribution from ${\cal O}_{V_2}$ ($C_{V_2} = +1.88$).]{
\includegraphics*[width=6.0cm]{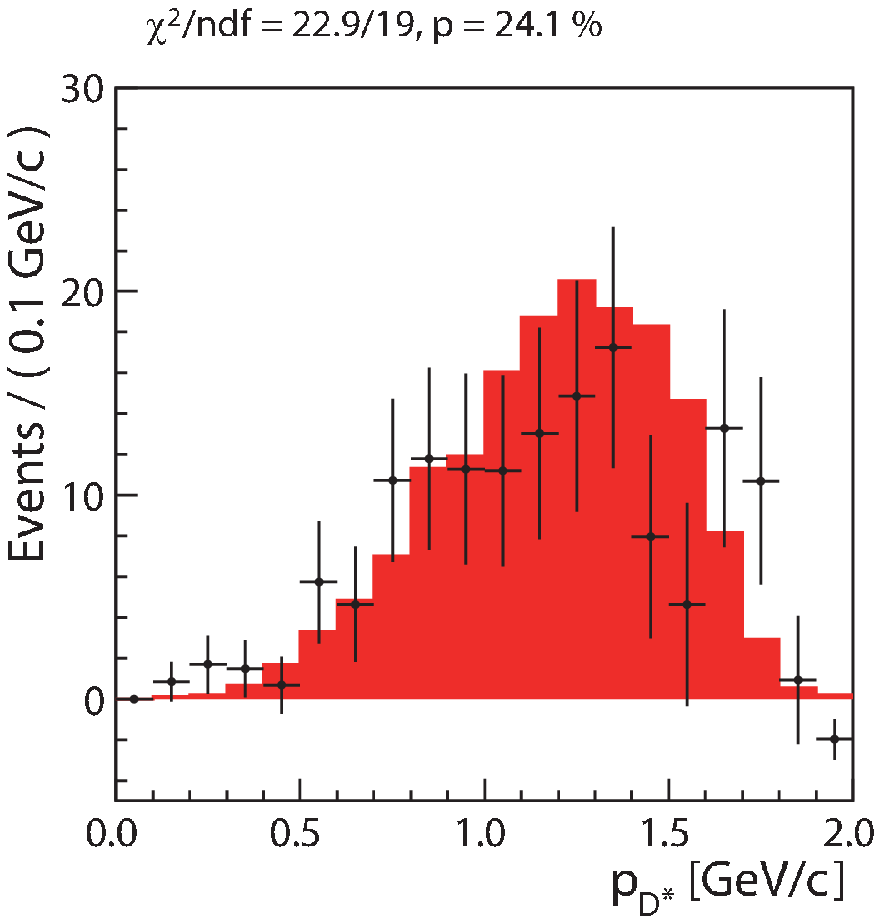}
\label{fig:kinematic_dstr_OPV2}}
\subfigure[SM with adding contribution from ${\cal O}_{T}$ ($C_{T} = +0.36$).]{
\includegraphics*[width=6.0cm]{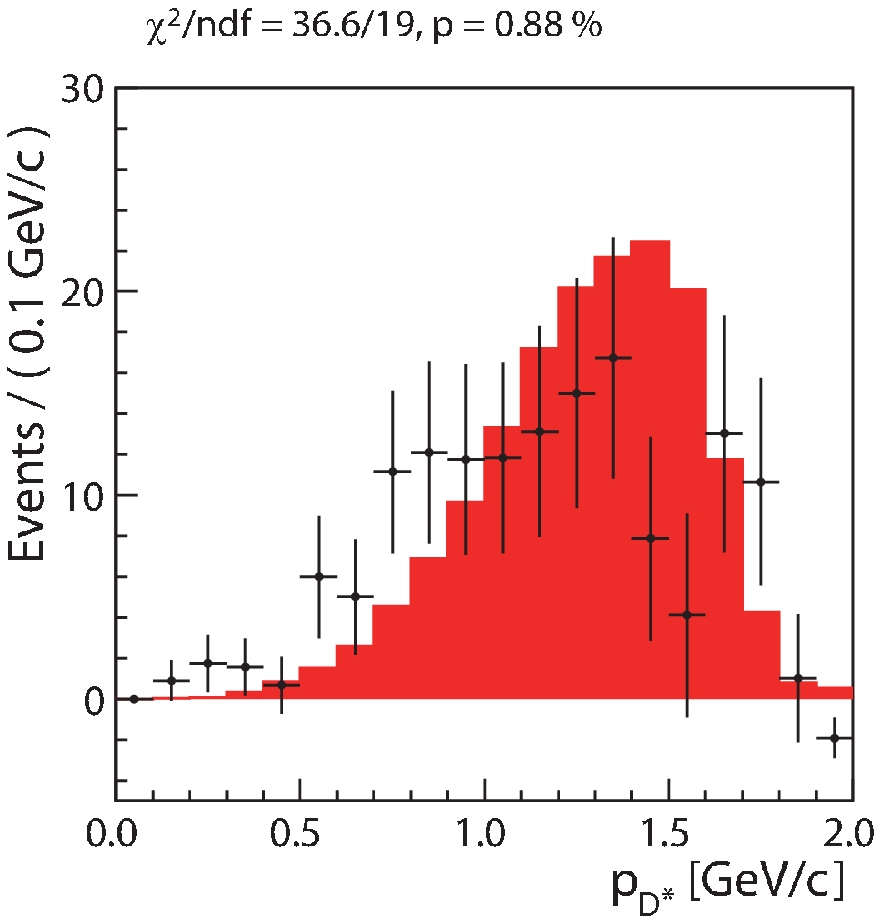}
\label{fig:kinematic_dstr_OPT}}
\subfigure[$R_2$-type leptoquark model with $C_{T} = +0.36$.]{
\includegraphics*[width=6.0cm]{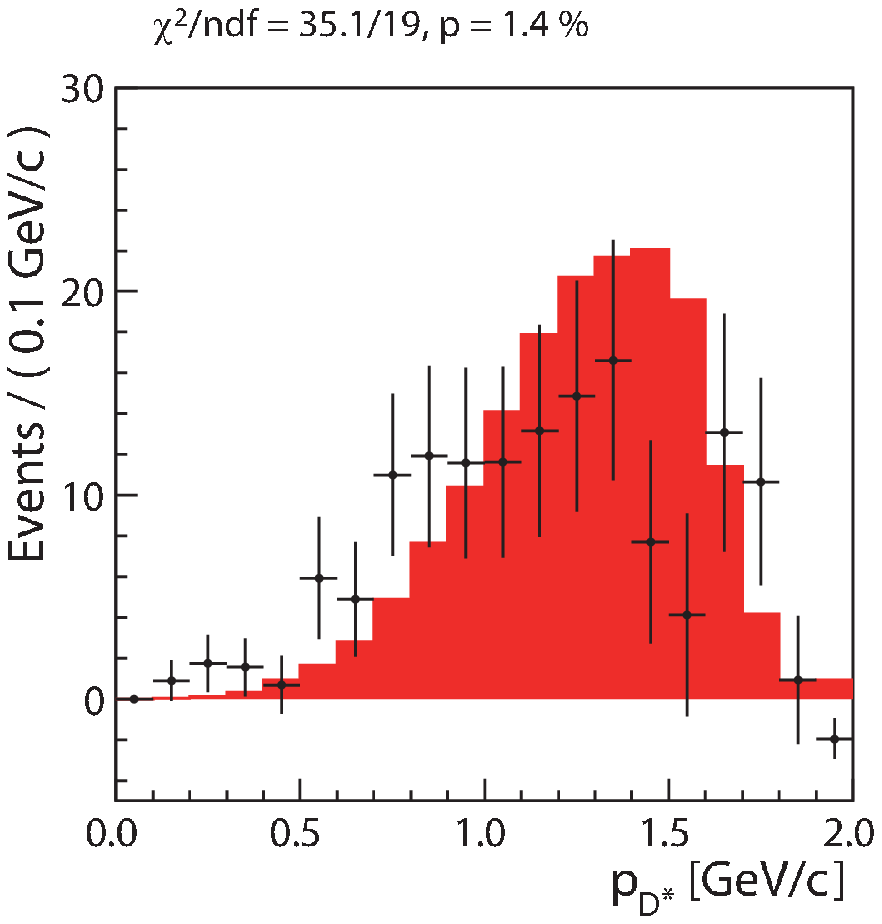}
\label{fig:kinematic_dstr_R2LQ}}
\subfigure[$S_1$-type leptoquark model with $C_{T} = +0.26$.]{
\includegraphics*[width=6.0cm]{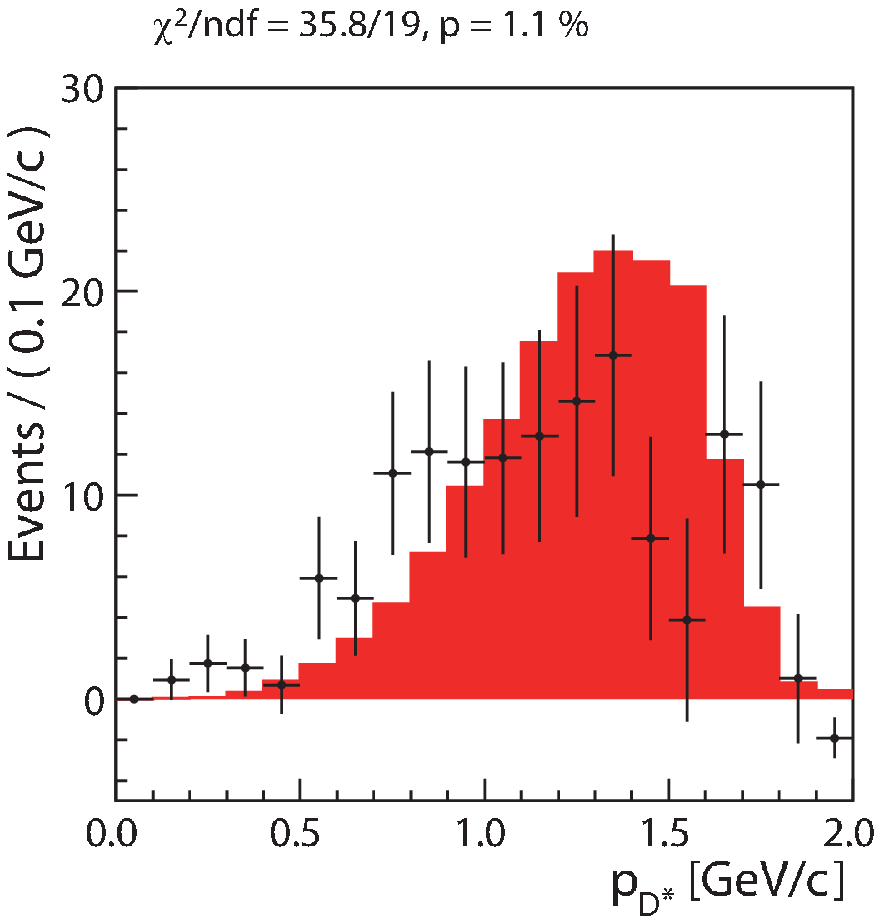}
\label{fig:kinematic_dstr_S1LQ}}
\caption{
Background-subtracted $D^*$ momentum distributions
in the region of ${\cal O}_{\mathit{NB}} > 0.8$ and $E_{\rm ECL} < 0.5$ GeV.
The points and the shaded histograms correspond to the measured and expected distributions, respectively.
The expected distributions are normalized to the number of detected events.
}
\label{fig:kinematic_dstr_supp}
\end{figure*}

\begin{figure*}[htb]
\centering
%\vspace{-25mm}
\subfigure[SM.]{
\includegraphics*[width=6.0cm]{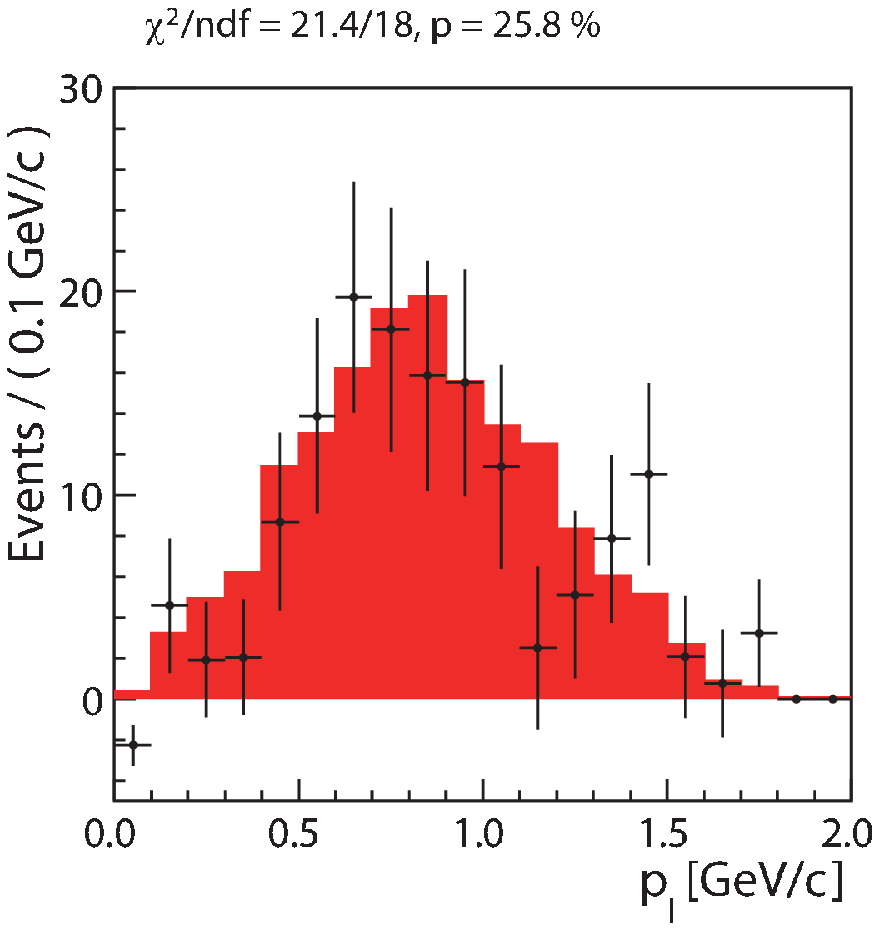}
\label{fig:kinematic_lepton_sm}}
\subfigure[Type-II 2HDM with $\tan \beta / m_{H^+} = 0.7$ GeV$^{-1}$.]{
\includegraphics*[width=6.0cm]{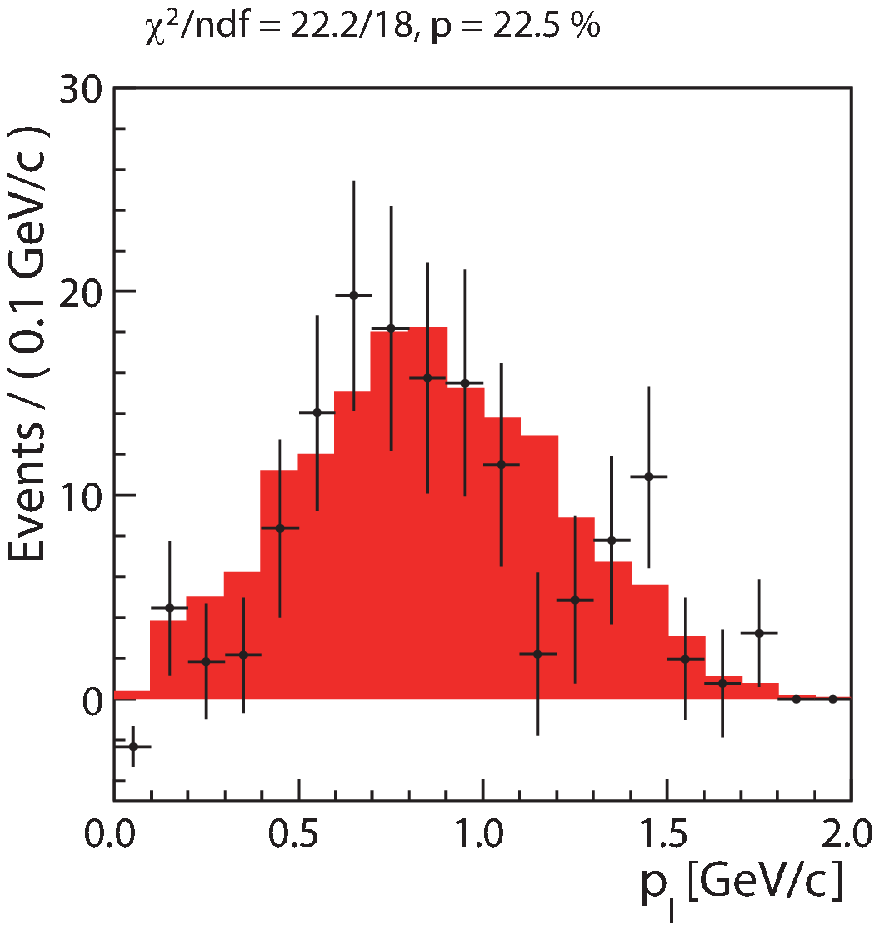}
\label{fig:kinematic_lepton_2hdm}}
%\subfigure[SM with adding contribution from ${\cal O}_{V_1}$ ($C_{V_1} = -2.10$).]{
%\includegraphics*[width=5.5cm]{kinematic_lepton_after_OPV1_m2p10.eps}
%\label{fig:kinematic_lepton_OPV1}}
\subfigure[SM with adding contribution from ${\cal O}_{V_2}$ ($C_{V_2} = +1.88$).]{
\includegraphics*[width=6.0cm]{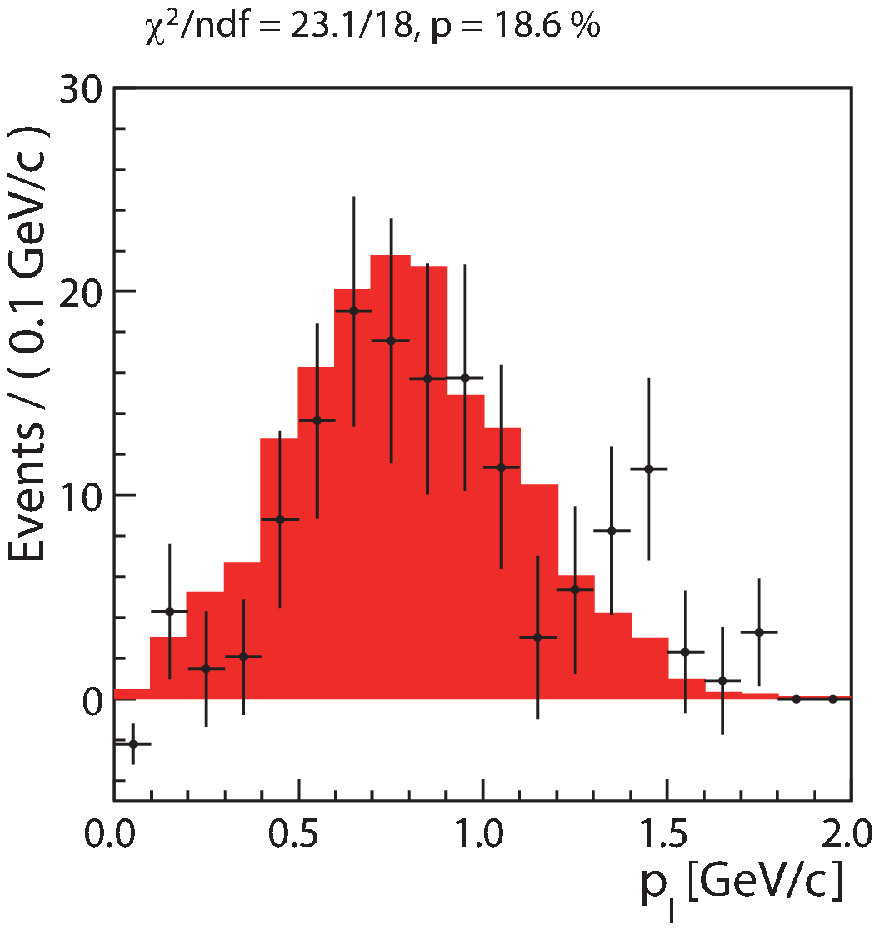}
\label{fig:kinematic_lepton_OPV2}}
\subfigure[SM with adding contribution from ${\cal O}_{T}$ ($C_{T} = +0.36$).]{
\includegraphics*[width=6.0cm]{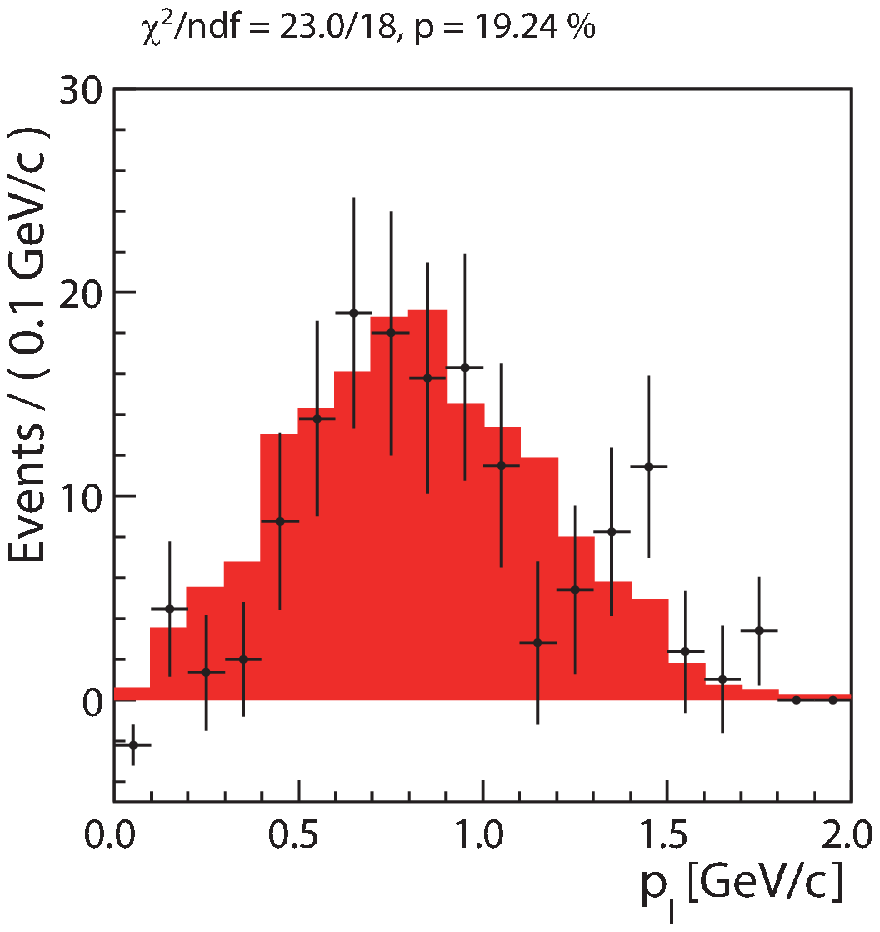}
\label{fig:kinematic_lepton_OPT}}
\subfigure[$R_2$-type leptoquark model with $C_{T} = +0.36$.]{
\includegraphics*[width=6.0cm]{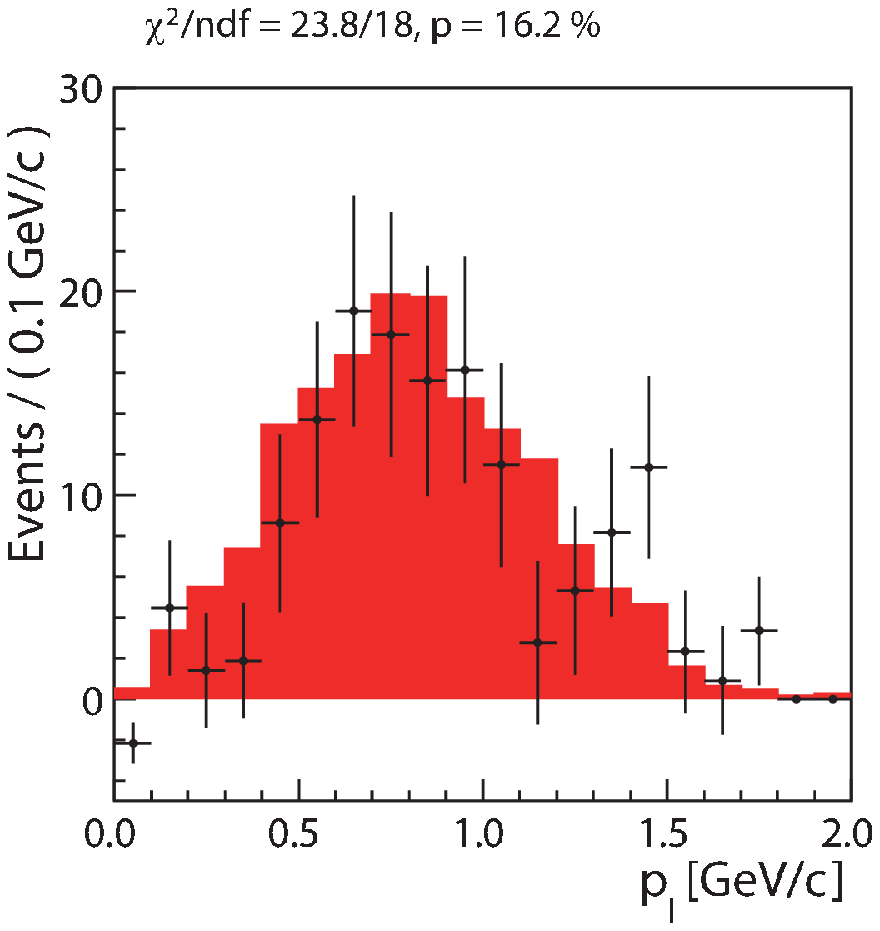}
\label{fig:kinematic_lepton_R2LQ}}
\subfigure[$S_1$-type leptoquark model with $C_{T} = +0.26$.]{
\includegraphics*[width=6.0cm]{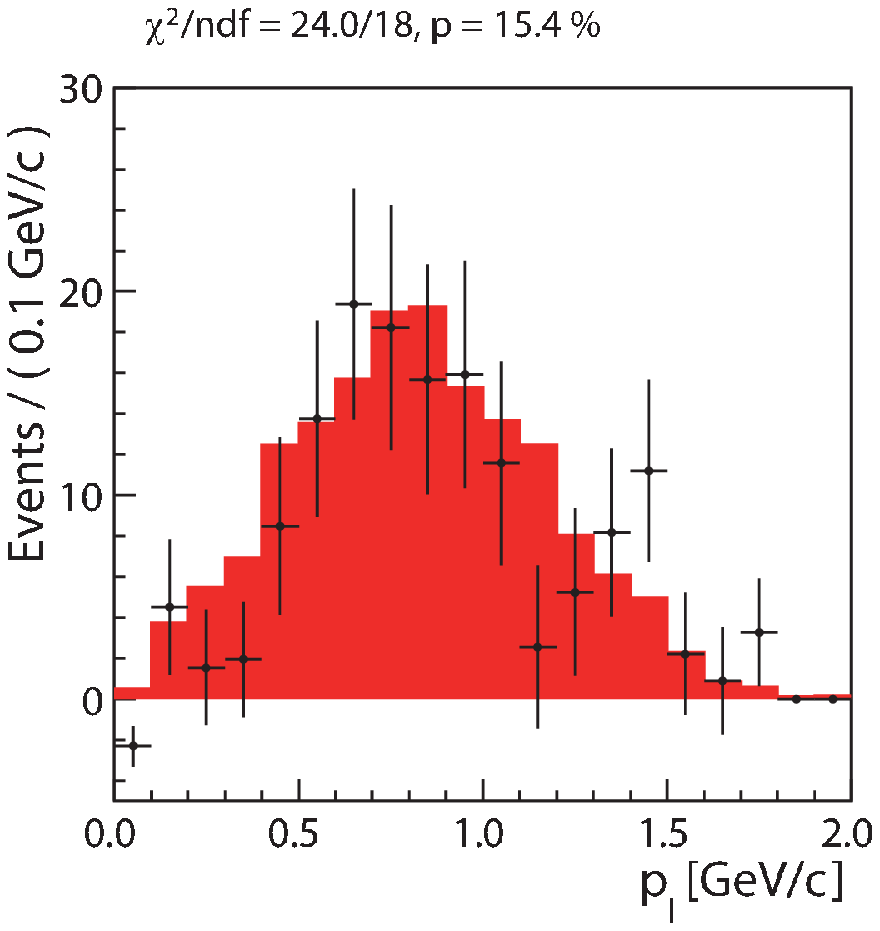}
\label{fig:kinematic_lepton_S1LQ}}
\caption{
Background-subtracted lepton momentum distributions
in the region of ${\cal O}_{\mathit{NB}} > 0.8$ and $E_{\rm ECL} < 0.5$ GeV.
The points and the shaded histograms correspond to the measured and expected distributions, respectively.
The expected distributions are normalized to the number of detected events.
}
\label{fig:kinematic_lepton_supp}
\end{figure*}

%\newpage


\begin{thebibliography}{99}
\bibitem{CHARGE_CONJUGATION}
Charge-conjugate decays are implied throughout this paper,
unless otherwise stated.
\end{thebibliography}
\end{document}